\documentclass[10pt]{article}
\pdfoutput=1
\usepackage{graphicx}
\usepackage{amsmath,amsfonts,amssymb} 
\usepackage{color}
\usepackage{cite}
\usepackage{slashed}
\usepackage{hyperref}

\newcommand{\unit}{\leavevmode\hbox{\small1\kern-3.6pt\normalsize1}}

\def\IC{\bf C}
\def\IZ{\bf Z}

\def\z2z2{$\IC^3/(\IZ_2\times\IZ_2)$}

\def\id{{\bf 1}}

\def\cW{\cal W}

\def\cp{\mbox{\bbbold C}\mbox{\bbbold P}}

\def\a{\alpha}

\def\b{\beta}

\def\d{\delta}\def\D{\Delta}

\def\k{\kappa}
\def\l{\lambda}

\def\p{\pi}

\def\s{\sigma}

\def\th{\theta}

\def\beq{\begin{equation}}\def\eeq{\end{equation}}
\def\beqa{\begin{eqnarray}}\def\eeqa{\end{eqnarray}}
\def\barr{\begin{array}}\def\earr{\end{array}}

\def\wt{\widetilde}

\def\ds {{\del \hspace{-6.4pt} \slash}\;}

 \let\br=\bigr

\def\bd{\begin{document}}
\def\ed{\end{document}}
\def\ba{\begin{array}}
\def\ea{\end{array}}
\def\bea{\begin{eqnarray}}
\def\eea{\end{eqnarray}}
\def\ft#1#2{{\textstyle{{\scriptstyle #1}\over {\scriptstyle #2}}}}
\def\fft#1#2{{#1 \over #2}}
\newcommand{\be}{\begin{equation}}
\newcommand{\ee}{\end{equation}}
\newcommand{\eq}[1]{(\ref{#1})}
\def\eqs#1#2{(\ref{#1}-\ref{#2})}
\def\det{{\rm det\,}}
\def\tr{{\rm tr}}
\newcommand{\ho}[1]{$\, ^{#1}$}
\newcommand{\hoch}[1]{$\, ^{#1}$}
\def\ra{\rightarrow}

\def\Xh{\hat{X}}
\def\ah{\hat{a}}
\def\xh{\hat{x}}
\def\yh{\hat{y}}
\def\ph{\hat{p}}
\def\G{{\cal G}}
\def\Dth{{\Delta_\th}}

\def\bk{{\bf k}}
\def\bx{{\bf x}}
\def\br{{\bf r}}
\def\tr{{\rm tr \,}}
\def\Tr{{\rm Tr \,}}
\def\diag{{\rm diag \,}}
\def\tg{{\rm tg \,}}
\def\NPB#1#2#3{Nucl. Phys. B {\bf #1} (19#2) #3}
\def\PLB#1#2#3{Phys. Lett. B {\bf #1} (19#2) #3}
\def\PLBold#1#2#3{Phys. Lett. {#1B} (19#2) #3}
\def\PRD#1#2#3{Phys. Rev. D {\bf #1} (19#2) #3}
\def\PRL#1#2#3{Phys. Rev. Lett. {\bf #1} (19#2) #3}
\def\PRT#1#2#3{Phys. Rep. {\bf #1} C (19#2) #3}
\def\MODA#1#2#3{Mod. Phys. Lett.  {\bf #1} (19#2) #3}
\def\ov{\overline}

\def\preal{{\rm Re\,}}
\def\pim{{\rm Im\,}}

\def\ds{\displaystyle}
\def\yzero{\smash{\hbox{$y\kern-4pt\raise1pt\hbox{${}^\circ$}$}}}
\def\p{\partial}
\def\a{\alpha}
\def\b{\beta}
\def\g{\gamma}
\def\d{\delta}
\def\beq{\begin{equation}}
\def\eeq{\end{equation}}
\def\beqa{\begin{eqnarray}}
\def\eeqa{\end{eqnarray}}
\def\Om{\Omega}
\def\om{\omega}
\def\th{\theta}
\def\vt{\vartheta}
\def\vphi{\varphi}
\def\-{\hphantom{-}}
\def\ov{\overline}
\def\s2{\frac{1}{\sqrt2}}
\def\wh{\widehat}
\def\wt{\widetilde}
\def\oh{\frac{1}{2}}
\def\tr{{\rm tr \,}}
\def\Tr{{\rm Tr \,}}
\def\diag{{\rm diag \,}}
\def\vac{|0 \rangle}
\def\vm{\relax{n_{\text{v}}}}
\def\cc{{\cal C}}
\def\ck{{\cal K}}
\def\ci{{\cal I}}
\def\cu{{\cal U}}
\def\cg{{\cal G}}
\def\cn{{\cal N}}
\def\cam{{\cal M}}
\def\cp{{\cal P}}
\def\ct{{\cal T}}
\def\cv{{\cal V}}
\def\cz{{\cal Z}}
\def\ch{{\cal H}}
\def\cf{{\cal F}}
\def\tv{\tilde v}
\def\Dsl{\,\raise.15ex\hbox{/}\mkern-13.5mu D} 
\def\IZ{Z\kern-.4em  Z}
\def\id{{\rm 1}}

\def\ti{\times}
\def\til{\tilde}
\def\eps{\epsilon}
\def\k{\kappa}
\def\A{\Arrowvert}
\def\cw{{\cal W}}
\def\G{\Gamma}
\def\car{{\cal R}}
\def\l{\lambda}
\def\raw{\rightarrow}
\def\Raw{\Rightarrow}
\def\inte{{\bf Z}}
\def\cpx{{\bf C}}
\def\real{{\bf R}}
\def\Lam{\Lambda}
\def\D{\Delta}
\def\cb{{\cal B}}
\def\ca{{\cal A}}

\def\lag{\mathcal{L}}
\def\sw{\sin\theta_W}
\def\cw{\cos\theta_W}

\def\sW{s_W}
\def\cW{c_W}
\def\sWh{s_{\hat{W}}}
\def\cWh{c_{\hat{W}}}

\def\mz{{M_Z}}
\def\mw{{M_W}}

\def\hZ{{\hat{Z}}}
\def\hW{{\hat{W}}}
\def\hX{{\hat{X}}}

\def\Zp{Z^{'}}

\def\gZpuV{g_{\Zp}^{u,V}}
\def\gZpdV{g_{\Zp}^{d,V}}
\def\gZpuA{g_{\Zp}^{u,A}}
\def\gZpdA{g_{\Zp}^{d,A}}

\def\mathZ{\mathcal{Z}}
\def\mathA{\mathcal{A}}
\def\M{\mathcal{M}}
\def\b#1{\bar{#1}}

\def\Tr{{\rm Tr}}

\def\p{\slashed{p}}

\def\Mnote#1{{\color{red}[VM: #1]}} 
\def\Pnote#1{{\color{blue}[MP: #1]}} 
\def\Snote#1{{\color{green}[PS: #1]}} 

\newcommand{\captions}{\sf\caption}

\def\lsim{\raise0.3ex\hbox{$\;<$\kern-0.75em\raise-1.1ex\hbox{$\sim\;$}}}
\def\gsim{\raise0.3ex\hbox{$\;>$\kern-0.75em\raise-1.1ex\hbox{$\sim\;$}}}

\renewcommand{\theequation}{\thesection.\arabic{equation}}
\allowdisplaybreaks

\begin{document}

\thispagestyle{empty}
\begin{flushright}
  FTUAM-15-5 ~~
  IFT-UAM/CSIC-15-018 ~~
 MAD-TH-15-03  \\

\end{flushright}

\begin{center}
  {\Large \textbf{Isospin violating dark matter in St\"uckelberg portal scenarios
  } }  
  
  \vspace{0.5cm}
  V\'ictor Mart\'in Lozano${}^{1,2}$,
  Miguel Peir\'o${}^{1,2}$,
  Pablo Soler${}^{3}$ \\[0.2cm] 
    
  {\textit{ 
      ${}^1$ 
      Instituto de F\'{\i}sica Te\'{o}rica UAM/CSIC, Universidad Aut\'{o}noma de Madrid,\\
      Cantoblanco, E-28049, Madrid, Spain\\[0pt]
      ${}^2$
      Departamento de F\'{\i}sica Te\'{o}rica, Universidad Aut\'{o}noma de Madrid,\\
      Cantoblanco, E-28049, Madrid, Spain\\[0pt]
      ${}^3$ 
      Department of Physics, 1150 University Avenue,\\
      University of Wisconsin, Madison, WI 53706, USA
  }}

\vspace*{0.7cm}

\begin{abstract}
Hidden sector scenarios in which dark matter (DM) interacts with the Standard Model matter fields through the exchange of massive $Z'$ bosons are well motivated by certain string theory constructions. In this work, we thoroughly study the phenomenological aspects of such scenarios and find that they present a clear and testable consequence for direct DM searches. We show that such string motivated St\"uckelberg portals naturally lead to isospin violating interactions of DM particles with nuclei. We find that the relations between the DM coupling to neutrons and protons for both, spin-independent ($f_n/f_p$) and spin-dependent ($a_n/a_p$) interactions, are very flexible depending on the charges of the quarks under the extra $U(1)$ gauge groups. We show that within this construction these ratios are generically different from $\pm1$ (i.e. different couplings to protons and neutrons) leading to a potentially measurable distinction from other popular portals. Finally, we incorporate bounds from searches for dijet and dilepton resonances at the LHC as well as LUX bounds on the elastic scattering of DM off nucleons to determine the experimentally allowed values of $f_n/f_p$ and $a_n/a_p$. 
\end{abstract}

\end{center}



\section{Introduction}

Uncovering the properties of Dark Matter (DM) and, in particular, its possible non gravitational interactions with visible matter is one of the greatest challenges of modern physics, and is accordingly the object of important experimental and theoretical efforts. 

A common theoretical framework for DM studies is the hidden sector scenario. In its minimal form, visible matter resides in a sector of the theory that hosts the Standard Model (SM) gauge and matter content (or simple extensions thereof), while DM resides in a hidden sector, with its own gauge and matter content, but is otherwise neutral under the SM group.

Within such a framework, several mechanisms have been proposed to mediate non-gravitational interactions between the different sectors, usually referred to as portals~\cite{Patt:2006fw,Feldman:2007wj,Falkowski:2009yz,Batell:2009di,Crivelli:2010bk,Chu:2011be,Essig:2013lka,Feng:2014cla,Feng:2014eja,Foot:2014uba,Bai:2014osa,Baek:2013dwa,Blum:2014jca,Cherry:2014xra,Arcadi:2014lta,Bian:2014cja}. Among them, perhaps the most popular is the Higgs portal~\cite{Patt:2006fw} in which the SM Higgs boson has renormalizable couplings to scalar fields of the hidden sector. This kind of construction leads to important phenomenological consequences such as the contribution of hidden final states to the branching fraction of the Higgs~\cite{Djouadi:2011aa,Djouadi:2012zc}. Another popular kind of portal is the $Z'$ portal. In this scenario, a hidden sector communicates with the SM via a gauge boson, provided that the SM is enlarged with an extra abelian gauge group~\cite{Langacker:2008yv}. The phenomenology of these constructions is very rich, and ranges from colliders to direct and indirect searches of DM~\cite{Dudas:2009uq,Cassel:2009pu,Frandsen:2011cg,Barger:2012ey,Arcadi:2013qia,Alves:2013tqa,Alves:2015pea}.

Particularly important for direct detection experiments are the isospin violation properties of the interactions of DM particles with nuclei induced by different portals. For example, the Higgs portal (at least in its simplest form) automatically predicts isospin preserving interactions. A similar rigidity in the pattern of isospin interactions is present in many other portals. Previous works have shown that a deviation from these patterns would require the presence of several mediators whose contributions to the cross section interfere, and hence, generate an amount of isospin violation potentially tunable~\cite{Gao:2011ka,Frandsen:2011cg,Belanger:2013tla,Hamaguchi:2014pja}.

It is the purpose of this work to show that, in contrast, $U(1)$ extensions of the SM with St\"uckelberg $Z'$ bosons acting as portals naturally accommodate rich patterns of isospin violating interactions. The latter, in turn, provide a clear and testable phenomenological consequence of such models. Extra abelian gauge factors are among the most common extensions of the SM~\cite{Langacker:2008yv}, and also among the best motivated from string theory, where massive extra $U(1)$ gauge bosons appear ubiquitously (for reviews see e.g. \cite{Ibanez:2012zz,Blumenhagen:2005mu,Blumenhagen:2006ci,Marchesano:2007de,Kakushadze:1997mc,Cleaver:1998gc}). In fact, when one tries to implement a visible sector with the SM gauge group in, say, intersecting brane models, one generically obtains not $SU(3)_c\times SU(2)_L\times U(1)_Y$, but rather
$U(3)_c\times U(2)_L\times U(1)^p$ which contains several extra abelian factors (including the centers of $U(3)_c$ and $U(2)_L$). 

The models we will consider along this work are based on this type of string constructions. The symmetry structure of this scenario can be represented schematically in the following form,
\begin{eqnarray}\label{stuckportal}
\!\!\!SU(3)_{c}\!\times\! SU(2)_{L}\!\times\!U(1)_{\text{v}}^n~\times~U(1)_{\text{h}}^m\!\times\! G_{\text{h}} \\[-10pt]
\!\!\!\underbrace{ \hphantom{SU(3)_{c}\!\times\! SU(2)_{L}\!\times\!U(1)_{\text{v}}^n}}_{\Psi_{\text{v}}} \hphantom{~\times~}\underbrace{ \hphantom{U(1)_{\text{h}}^m\!\times\!G_{\text{h}}}}_{\Psi_{\text{h}}}\nonumber
\end{eqnarray}
where the $U(1)_{\text{v}}^n$ are $n$ abelian gauge factors to which the visible matter fields $\Psi_{\text{v}}$ couple. All of the corresponding gauge bosons acquire a mass through the St\"uckelberg mechanism, except for a particular linear combination of them that corresponds to hypercharge and remains massless (in the phase of unbroken electroweak symmetry). $U(1)_{\text{h}}^m$ are $m$ abelian gauge factors (some of which could be massless) to which only hidden matter $\Psi_{\text{h}}$ couples, and $G_{\text{h}}$ represents the semi-simple part of the hidden gauge group.

\begin{figure}[tbp]
\centering
\includegraphics[width=250pt]{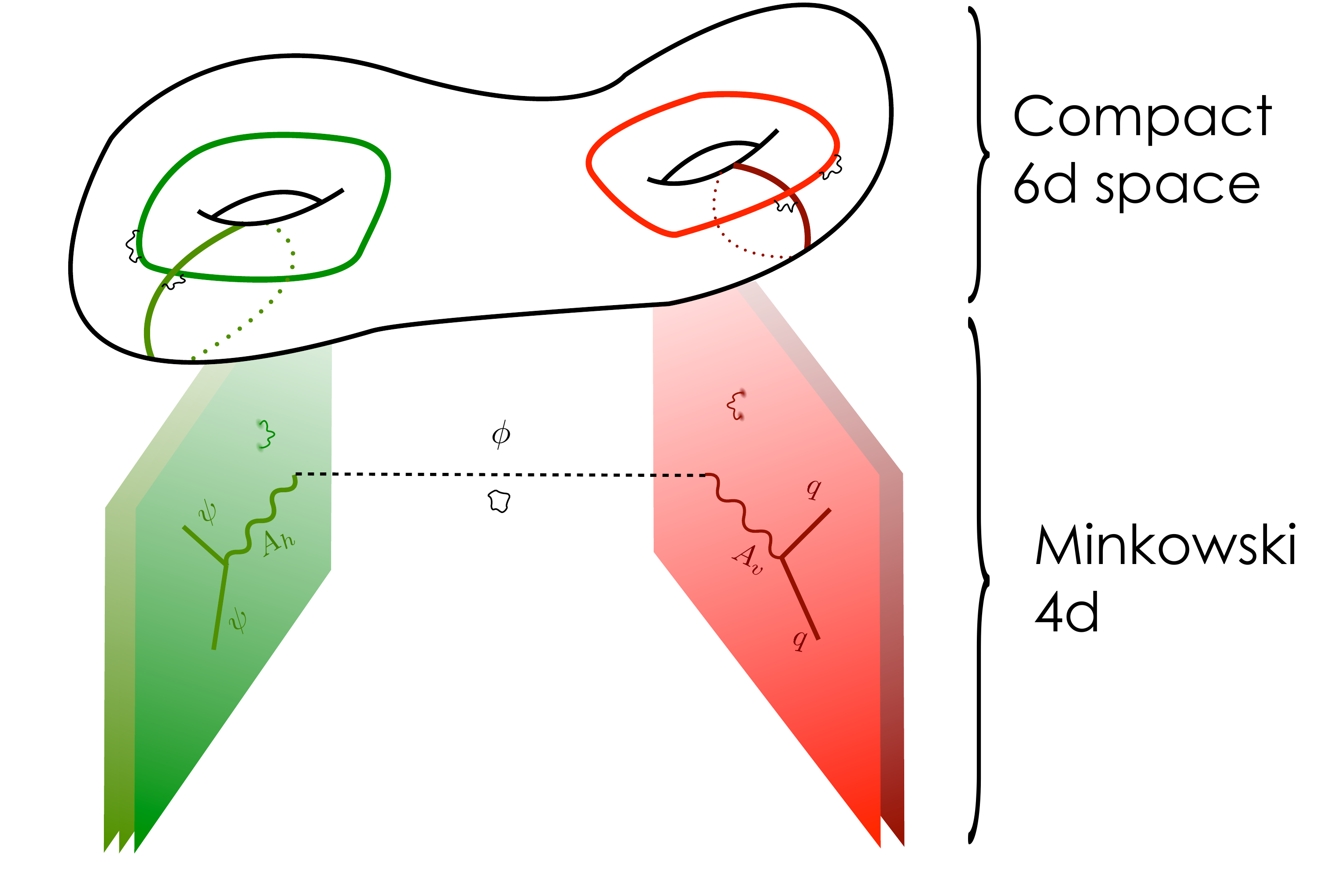}
\hspace{2cm}
\includegraphics[width=100pt]{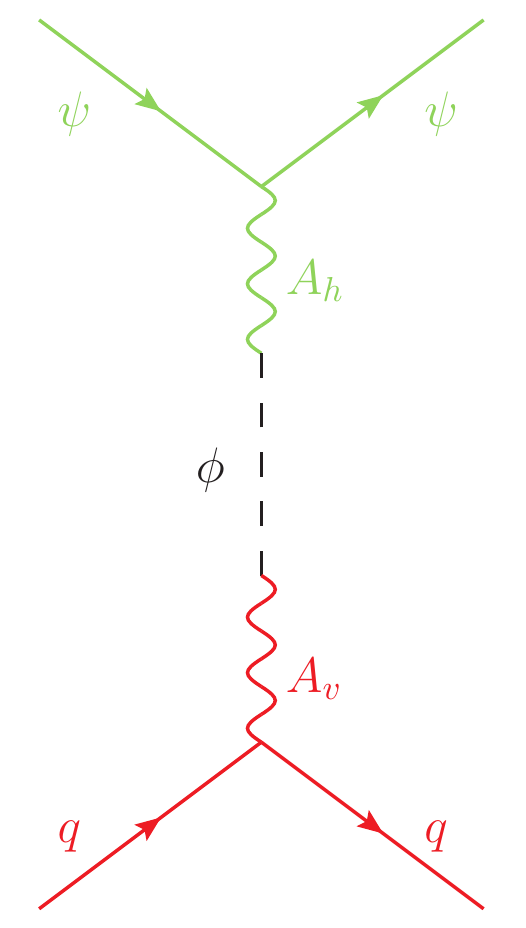}
\caption{\small Left: Schematic representation of a hidden sector scenario~\eqref{stuckportal} with intersecting D-branes. Green and red branes do not intersect each other and hence host different sectors. Right: Diagram contributing to the elastic scattering of hidden sector DM, $\psi$, off quarks. The mediator of this interaction is a mixing of different string axions, $\phi$, and the vectors $A_v$ and $A_h$.}\label{drawing}
\end{figure}

As mentioned, these type of scenarios can fairly easily be implemented in models of intersecting D6-brane of type IIA string theory. Intuitively, each sector consists of several intersecting stacks of branes wrapping 3-cycles of a six-dimensional compactification space. Each stack hosts a $U(N)$ gauge factor and chiral matter arises at the brane intersections. Different sectors arise from brane stacks that do not intersect each other and can hence be separated in the internal space (see Fig.~\ref{drawing}).

The extra abelian gauge bosons of Eq.~\eqref{stuckportal} can provide a portal into the hidden sector in two different ways. The most thoroughly studied is a small kinetic mixing of a light hidden gauge boson with the visible massless photon~\cite{Holdom:1985ag,Lust:2003ky,Abel:2003ue,Abel:2006qt,Abel:2008ai,Goodsell:2009xc,Cicoli:2011yh,Gmeiner:2009fb,Honecker:2011sm}. This generates an effective coupling of DM with visible fields which is proportional to their electric charge, and hence, the DM particles only couple to protons and do not couple to neutrons. From the point of view of direct DM searches, this is very important, since the elastic scattering of DM off nucleons only receives contributions from protons. This is, the ratio between the coupling of DM to neutrons and protons vanishes, $f_n/f_p=0$.

The second mechanism, which will be the main subject of this paper, was pointed out in Refs.~\cite{Feng:2014eja,Feng:2014cla} (see also~\cite{Kors:2004dx,Dermisek:2007qi,Verlinde:2007qk,Shiu:2013wxa}).
It results from the mixing of massive $U(1)$s of the visible sector with $U(1)$s of the hidden one. Despite living in different sectors, the $U(1)$ gauge bosons $A_{\text{v}}^n$ and $A_{\text{h}}^m$ often have St\"uckelberg couplings to the same axions, e.g. RR closed string axions in type II string models. As a consequence the resulting mass matrix can be highly non-diagonal. The `physical' $Z'$ eigenstates obtained  upon diagonalization of the kinetic and mass matrices are largely mixed combinations of $A_{\text{v}}^n$ and $A_{\text{h}}^m$ and hence couple simultaneously to both, visible and hidden, matter sectors. This mass mixing is a tree-level effect that provides an effective portal into hidden sectors, provided the 
associated $Z'$ bosons are light enough. 

Despite the potentially complex gauge and matter structure of the hidden sector, it seems reasonable to assume that it hosts a Dirac fermion $\psi$ that plays the role of DM in the Universe. The stability of these particles is easily guaranteed by the perturbatively conserved $U(1)_h^m$ symmetries or by non-perturbatively exact discrete subgroups thereof, or simply because they are the lightest particles of the whole sector. In any case, their interaction with the SM fermions will be driven by the exchange of a $Z'$ boson. 
For DM direct detection experiments the leading interaction of the elastic scattering of $\psi$ with quarks is depicted in the right panel of Fig.~\ref{drawing}. Following this reasoning, it is clear that the charges of the SM fermions under the $U(1)^n_{\text v}$ groups that mix with the hidden sector will determine the prospects for detecting $\psi$ in these experiments. 

In this work we study the phenomenology of a class of scenarios of this kind that can be embeded into well known string theory constructions. In particular, we focus on the isospin violation character of the DM interactions with protons and neutrons induced by the $Z'$ bosons. As we will see, isospin violation could distinguish these St\"uckelberg portal models from other popular setups, such as the Higgs portal or the $Z$-mediation scenarios.

This work is organized as follows. In section~\ref{sec:setup} we review the general theoretical framework that underlies our models. We describe the mixing mechanism that generates an off-diagonal mass matrix for the $U(1)$ gauge bosons and study general properties the eigenstates of such matrix, which are the physical $Z'$ fields that communicate the hidden and the visible sectors. We also discuss how the general form of the effective Lagrangian arises from certain string compactifications with intersecting D6-branes. In section~\ref{sec:SMcouplings} we take a well known class of such string models and determine the SM couplings to the lightest $Z'$ mediator in terms of a few mixing parameters. In section~\ref{sec:isospinviolation} we study the isospin violation properties of the DM-nucleon interactions in these constructions in terms of these parameters, and compare them to those arising in other popular scenarios. In section~\ref{sec:bounds} we incorporate to our analysis bounds from direct detection (LUX) and collider searches (LHC)
for six benchmark points in the parameter space of the model. Finally, we give some concluding remarks in section~\ref{sec:conclusions}.

\section{Effective Lagrangian and $Z'$ eigenstates}\label{sec:setup}

In this section we review the general constructions of Refs.~\cite{Feng:2014eja,Feng:2014cla} which describe the mixing mechanism of massive $U(1)$ gauge bosons from different sectors, the so-called St\"uckelberg portal. We begin with a discussion in terms of the effective field theory, and describe later on the string implementation of such setup.

\subsection{Non-diagonal $U(1)$ mass matrix}\label{subsec:massmatrix}
The abelian sector of the construction sketched in Eq.~\eqref{stuckportal} can generically be described by the Lagrangian
\begin{equation}\label{lag}
{\cal L}=-\frac{1}{4}\vec{F}^{\,T}\cdot f\cdot\vec{F}-\frac{1}{2} \vec{A}^{\,T}\cdot M^2\cdot\vec{A}+
\sum_\alpha\overline{\psi}_\alpha \left(i\partial\!\!\!/ +  \vec{Q}_{\alpha}^{\,T}\cdot\vec{A}\!\!\!/ \right)\psi_\alpha
\end{equation}
where the vector $\vec{A}^{\,T}=(A_1\ldots  A_{n+m})$ encodes all the $U(1)$ gauge bosons of the system, with field strength $\vec{F}=d\vec{A}$.  In this normalization the gauge coupling constants are absorbed in the kinetic matrix $f$. In hidden sector scenarios, the charge vectors $\vec{Q}_\alpha$ of a given matter field $\psi_\alpha$ will have non-zero entries only for one of the sectors (either visible or hidden), while the kinetic and mass matrices $f$ and $M$ can have off-diagonal entries that mix both sectors. We are interested in particular in the mixings induced by the mass matrix $M$. 

The mass terms for Abelian gauge bosons $\vec{A}$ can be generated by either the Higgs or the St\"uckelberg mechanisms. In both cases the crucial term in the Lagrangian is the coupling of $\vec{A}$ to a set of pseudo-scalar periodic fields $\phi^i\sim \phi^i+2\pi$ whose covariant kinetic terms read
\begin{equation}\label{St-lag}
{\cal L}_M =-\frac{1}{2} \, G_{ij}\left(\partial\phi^i - k_a^i A^a\right)\left(\partial\phi^i - k_b^j A^b\right)\,.
\end{equation}
Here, $G_{ij}$ corresponds to a positive-definite kinetic matrix (the metric in the space of $\phi^i$ fields), which in our conventions has dimension of $({\rm mass})^2$. The factors $k_a^i$ encode the non-linear gauge transformations 
\begin{equation}
\vec{A}\to\vec{A}+ d\vec\Lambda \qquad\Longrightarrow \qquad \phi^i \to \phi^i + k^i_a \Lambda^a\,.
\end{equation}
The statement that the gauge symmetry group is compact ($U(1)$ rather than $\mathbb R$) implies that the transformations must be periodic $\Lambda\sim\Lambda +2\pi$, and hence that the $k_a^i$ factors (as well as the matter charges $\vec{Q}_\alpha$ of Eq.~\eqref{lag}) must be quantized. In fact, under the appropriate normalization they can be assumed to be integers, $k_a^i,\,Q_\alpha^a\subset \mathbb Z$. In the case of the Higgs mechanism, the axion-like fields $\phi^i$ are identified with the phases of Higgs fields $H^i$, and the $k_a^i$ factors are simply the charges of the latter under the $U(1)_a$ groups. It is not surprising that these are integer quantities. What is perhaps less obvious is that even for axions not related to a Higgs fields, one can still associate $U(1)$ integer ``charges'' that determine their gauge transformations.

The $U(1)$ gauge bosons get a mass by absorbing the axions, $\phi^i$, through the Lagrangian~\eqref{St-lag}. After gauge fixing the $U(1)$ symmetries, the mass term of equation~\eqref{lag} is generated, and the corresponding mass matrix takes the form
\begin{equation}
M^2=K^T\cdot G\cdot K\,.
\end{equation}
It is easy to see that this matrix can be highly non-diagonal and have off-diagonal entries that mix hidden and visible sectors. This can happen with particular strength if the mixing is induced by the integer matrix $K$ of axionic charges.

 The dynamical origin of the mixing is the simultaneous coupling of vector bosons from different sectors to the same axions (see figure~\ref{drawing}). As a toy model, consider two $U(1)$ gauge bosons, a visible $A_v$ and a hidden $A_h$, that couple to an axion with charges +1 and -1, respectively, i.e. $K=\binom{+1}{-1}$, whose kinetic matrix is $G=m^2$. The resulting mass matrix of the $U(1)$ bosons would read $M^2=m^2 \left(\begin{smallmatrix} 1&-1\\-1&1\end{smallmatrix}\right)$. The resulting physical eigenstates are obviously highly mixed combinations of $A_v$ and $A_h$ that hence couple with similar strength to both sectors.
 
In the following we generalize this simple example to the case where several gauge bosons mix with each other by absorbing several axions with mixed charges. 
 
\subsection{Diagonalization and eigenstates}\label{subsec:eigenstates}

In order to study the properties of the system described by the Lagrangian of Eq.~\eqref{lag}, it is convenient to move to a basis in which the gauge bosons have a canonical kinetic term and a diagonal mass matrix. The former can be obtained by a linear transformation:
 \begin{equation}
 \vec{A}\equiv \Lambda \cdot \vec{A'}
 \end{equation}
such that $\Lambda^{\,T}\cdot f\cdot \Lambda = 1$. In the case with no kinetic mixing, i.e. $f=\mathrm {diag}(g_1^{-2},\ldots, g_N^{-2})$, the transformation matrix is simply $\Lambda=\mathrm {diag}(g_1,\ldots, g_N)$. For the moment we need not assume such simplification, and we work with a general kinetic matrix $f$.

The Lagrangian in terms of the transformed bosons $\vec{A}'$ reads
\begin{equation}\label{lag2}
{\cal L}=-\frac{1}{4}\vec{F'}^2-\frac{1}{2} \vec{A'}^{\,T}\cdot \Lambda^T \cdot M^2\cdot\Lambda\cdot\vec{A'}+\,\overline{\psi}\left(i\partial\!\!\!/ +  \vec{Q}_{\psi}^{\,T}\cdot\Lambda\cdot\vec{A}\!\!\!/\, '\right)\psi\,.
\end{equation}
Notice that with this new normalization, what appears in the matter coupling to the gauge boson is no longer just the charges $Q$, but products of these and coupling constants $g$ (and possible kinetic mixing parameters). 

We need now an orthogonal transformation ${\cal O}$ that diagonalizes the mass matrix $\tilde{M}^2\equiv\Lambda^T \cdot M^2\cdot\Lambda$. That is, we need to find a basis of orthonormal eigenvectors
\begin{equation}\label{Ort}
\tilde{M}^2\cdot\vec{v_i}=m_i^2 \vec{v_i}\qquad \Longrightarrow\qquad {\cal O}=\left( \vec{v_1}\, \,\vec{v_2}\,\,\ldots \,\vec{v_N}\right)\,.
\end{equation}
Conveniently, we define $\vec{v_i}'\equiv \Lambda\cdot \vec{v_i}$. 
The transformation $\vec{A}'\equiv{\cal O}\cdot\vec{A}''$ brings the Lagrangian to a standard form with canonical kinetic term and diagonal mass matrix:
\begin{equation}\label{physlag}
{\cal L}=-\frac{1}{4} {F''}_i^2 -\frac{1}{2} m_i^2 {A''}_i^2+
\sum_\alpha\overline{\psi}_\alpha \left(i\partial\!\!\!/ +  \vec{g'}_{\alpha}^{\,T}\cdot\vec{A''}\!\!\!\!\!/ \,\right)\psi_\alpha
\end{equation}
The coupling of a vector $A''_i$ to the matter field $\psi_\alpha$ is given by a linear combination of the original charges:
\begin{equation}\label{couplings}
{g'}_\alpha^{(i)}=\vec{Q}_\alpha^{\,T}\cdot \vec{v}_{\,(i)}'\,.
\end{equation}

Notice the important fact that, for massless eigenvectors, $\vec{v_i}'$ are precisely the zero eigenvectors of the original mass matrix $M^2$, i.e. they satisfy $K\cdot \vec{v_i}'=0$. Since the entries of the matrix $K$ are integer numbers, the entries of the massless eigenvectors $\vec{v_i}'$ will be also integers, up to an overall normalization factor. The corresponding gauge bosons will be massless, have quantized charges, and if the form of the matrix $K$ is appropriate, will couple exclusively to one sector of the theory. They are hence perfect candidates to play the role of the SM hypercharge.

These last remarks do not apply to massive eigenstates, for which $M^2\cdot \vec{v_i}'\neq \alpha_i \vec{v_i}'$. Generically, given the non-diagonal character of the mass matrix $M^2$, all of the entries of the massive eigenvectors $\vec{v_i}$ will be non-zero and of the same order. The physical massive gauge bosons $A''_i$ will be hence a linear combination of both visible and hidden bosons and they will act as portals into hidden sectors.

Before concluding this section let us write down an important condition on the vectors $\vec{v_i}$. The orthogonality of the transformation matrix $\cal O$ of Eq.~\eqref{Ort} translates into the condition 
\begin{equation}\label{orthogonality}
\vec{v_i}^{\,T}\cdot\vec{v_j}=\delta_{ij}\qquad \Longrightarrow \qquad \vec{v_i}'^{\,T}\cdot f\cdot \vec{v_j}'=\delta_{ij}\,.
\end{equation}
We will have to take this condition into account in the phenomenological analysis carried out in the following sections.

\subsection{The string theory interpretation}\label{subsec:stringtheory}

As mentioned in the introduction, one nice feature of the St\"uckelberg portal is that it finds a natural implementation in string theory, and a particularly intuitive one in models of intersecting D-branes. A detailed study and explicit examples in the setup of toroidal orientifolds of type IIA string theory can be found in the original references~\cite{Feng:2014eja,Feng:2014cla}. Here, we briefly describe where the different fields and couplings arise in such models (for general reviews on these type of string compactifications, see e.g.~\cite{Ibanez:2012zz,Blumenhagen:2005mu,Blumenhagen:2006ci,Marchesano:2007de}). 

In type IIA orientifold compactifications, gauge bosons arise from open strings living on D6-branes that span the four non-compact dimensions, and wrap three-cycles of the six dimensional compactification space $X_6$ (usually a Calabi-Yau manifold). A stack of $N$ overlapping such branes usually hosts a gauge group $U(N)\cong SU(N)\times U(1)$. Chiral matter fields arise at the intersections of two stacks. Hence, in order to obtain hidden sector scenarios, one has to choose carefully the cycles wrapped by the branes to make sure that stacks from different sectors do not intersect with each other.

The abelian gauge bosons living in such stacks couple not only to open strings, but also to closed strings which include the graviton, and also Ramond-Ramond (RR) axions that arise from the reduction of RR three forms along three-cycles of $X_6$. Being associated to closed strings that propagate in the bulk of the compactification, it is natural to consider that such RR axions couple to gauge fields from different sectors. These couplings are of the St\"uckelberg type given in Eq.~\eqref{St-lag} and generate masses for the gauge bosons. The charge matrix $K$ is determined by the wrapping numbers of the branes around odd cycles of $X_6$ (odd with respect to the orientifold projection), and can be engineered in such a way that the mass matrix is highly non-diagonal.

The matrix $G$ that also enters in the formula for the mass matrix $M^2$ and is identified with the complex structure moduli space metric of the compactification space $X_6$, times a string scale factor $M_s^2$. Unfortunately, except for the simplest compactifications, this metric is unknown. Nevertheless, as long as some RR axion has non-zero charges under $U(1)$ groups from different sectors, the mixing induced by the mass matrix $M^2$ is expected to be strong and results in physical $Z'$ bosons that couple visible and hidden sectors.

The final ingredient in the Lagrangian of Eq.~\eqref{lag} is the kinetic matrix $f$. At tree level, this matrix is diagonal $f=\mathrm {diag}(g_1^{-2},\ldots, g_N^{-2})$, with the couplings determined by the volume of the cycles wrapped by the corresponding branes. Loop corrections can generate off-diagonal terms that produce small kinetic mixings among different $U(1)$s.

The fate of the $U(1)$ gauge bosons in this type of models is to gain a mass of the order of the string scale, suppressed by the square of the gauge coupling factor, $m_{Z'}\propto g^2 M_s^2$. This is expected to be very large in a broad class of string constructions. Nevertheless, several mechanisms have been proposed to lower the $Z'$ masses, including large volume and anisotropic compactifications, or eigenvalue repulsion effects~\cite{Ghilencea:2002da,Goodsell:2009xc,Feng:2014cla}. The conclusion is that, although not generic, $Z'$ masses at scales as low as the TeV, or even smaller, can be achieved in several setups.

At energy scales much lower than the $Z'$ boson masses, the corresponding $U(1)$ symmetries become effectively global. They are in fact perturbatively exact symmetries of the effective Lagrangian, and they are broken only by highly suppressed non-perturbative effects~\cite{Blumenhagen:2006xt,Ibanez:2006da,Florea:2006si}. Therefore, the $U(1)$ symmetries that extend the visible sector gauge group in realistic D-brane constructions should find an interpretation in terms of known approximate global symmetries of the SM, such as Baryon or Lepton number.

Interestingly, these extra $U(1)$ groups are generically anomalous symmetries of the SM. It is well known, however, that these anomalies are cancelled by a generalized Green-Schwarz mechanism, in which the RR axions, $\phi^i$, and the St\"uckelberg couplings of Eq.~\eqref{St-lag} play a crucial role. Although the gauge bosons associated to such anomalous $U(1)$s are not considered too frequently in the phenomenological literature, they are a key (and in fact most often unavoidable) ingredient of realistic constructions with open strings. 

In the following, we take the St\"uckelberg portal string constructions we have described in this section as a motivation, and study some of their phenomenological consequences.

\section{SM fermion couplings to $Z'$}\label{sec:SMcouplings}

As in Refs.~\cite{Feng:2014eja,Feng:2014cla}, we focus on visible sectors realized as in~\cite{Ibanez:2001nd}, the so-called {\it Madrid quivers}, which provide some of the simplest realistic models of intersecting D6-branes. In order to reproduce the SM one introduces four stacks of branes yielding a $U(3)_A\times U(2)_B \times U(1)_C\times U(1)_D$  visible gauge group. 
The intersection numbers of these branes are chosen in such a way that the model reproduces the SM chiral spectrum and is free of anomalies (with anomalies of extra $U(1)$ factors cancelled by the Green-Schwarz mechanism). 

In Table~\ref{tab:u1} the charges of the SM particles under the four visible $U(1)$ factors are presented. These charges can be interpreted in terms of known global symmetries of the SM. In particular, $Q_A$ and $Q_D$  are proportional to baryon and lepton number, respectively. 
With these charge assignments, the hypercharge corresponds to the linear combination
\begin{equation}
Q^Y=\frac{1}{6}\left(Q_A-3Q_C+3Q_D\right)\,.
\label{eq:hypercharge}
\end{equation}
One has to make sure that such a combination remains as a massless gauge symmetry of the system (before electroweak symmetry breaking), i.e. that it corresponds to a zero eigenstate of the mass matrix $M^2$. Following the discussion below Eq.~\eqref{couplings}, one has to make sure hence that the matrix of axionic charges $K$ has an eigenvector $\vec{v_Y}'=(1,0,-3,3;0,\ldots,0)$ with zero eigenvalue. The first entries of this vector correspond to the visible sector, and the latter to the hidden one, so that hypercharge couples exclusively to visible matter.
In fact, this condition can be implemented in type II string constructiones by simple topological requirements on the wrapping numbers of the visible branes. Therefore, according to Eq.~\eqref{couplings}, the hypercharge coupling to a matter field $\psi_\alpha$ reads
\begin{equation}
g_{\alpha}^Y=e\,Q_{\alpha}^Y=\frac{e}{6}\left(Q_{\alpha A}-3Q_{\alpha C}+3Q_{\alpha D}\right)\,.
\end{equation}

In general, the remaining three visible $U(1)$ gauge bosons acquire masses by the St\"uckelberg mechanism, and as stressed in the previous section, they can have strong mass mixing with hidden $U(1)$ bosons. In this work we are interested in the phenomenology induced by the lightest of the resulting physical $Z'$ bosons whose contribution to the DM interaction with SM particles is dominant. Since we cannot know the explicit form of the mass matrix $M^2$ for generic string compactifications (in particular because of the lack of control of the $G$ matrix that enters the Lagrangian) we will simply parametrise the couplings of the lightest $Z'$ boson to the matter fields $\psi_\alpha$ by a linear combination
\begin{equation}
g_{\alpha}^{Z'}= a \,Q_{\alpha A}+b\, Q_{\alpha B}+c\, Q_{\alpha C}+d\, Q_{\alpha D} + \sum_{i=1}^m h_i\,Q_{\alpha i}^{(h)}
\label{eq:charges}
\end{equation}
where we have included the contributions from hidden $U(1)$ factors. The parameters $a,\,b,\,c,\,d$ and $h_i$ are precisely the entries of the vector $\vec{v\,}'_{Z'}=(a,b,c,d;h_1,\ldots)$ of Eq.~\eqref{couplings}. For massive $Z'$ bosons these are continuous parameters and as already stressed, they are all generically different from zero. Furthermore, notice that, by definition, $\vec{v\,}'_{Z'}\equiv \Lambda\cdot \vec{v}_{Z'}$, where $\vec{v}_{Z'}$ is a normalized vector. Since at tree level $\Lambda=\mathrm {diag}(g_1,\ldots, g_N)$, one can see that the parameters $a,\,b,\,c,\,d$ and $h_i$ will be proportional to the original gauge coupling constants, and hence perturbative. 

\begin{table}
\begin{center}
    \begin{tabular}{|c|c|c|c|c|c|}
    \hline
    Matter field & $Q_A$& $Q_B$& $Q_C$& $Q_D$&$Y$\\[3pt]
    \hline
    $Q_L$&1&-1&0&0&1/6\\[3pt]
    $q_L$&1&1&0&0&1/6\\[3pt]
    $U_R$&-1&0&1&0&-2/3\\[3pt]
    $D_R$&-1&0&-1&0&1/3\\[3pt]
    \hline
    $L$ &0&-1&0&-1&-1/2\\[3pt]
    $E_R$ &0&0&-1&1&1\\[3pt]
    $N_R$ &0&0&1&1&0\\[3pt]
        \hline
    \end{tabular}
          \caption{\footnotesize
 SM spectrum and $U(1)_i$ charges in the four stack models of Ref.~\cite{Ibanez:2001nd}. Anomaly cancellation requires the three quark families to be divided into two $Q_L$ doublets and  two antidoublets $q_L$ of $U(2)_B$ i.e. they differ in their $U(1)_B$ charge. We assign the up and down quarks to the antidoublets.  
}
      \label{tab:u1}
\end{center}
\end{table}

The parameters $a$, $b$, $c$ and $d$, which in turn determine the effective couplings of visible matter to DM, are nevertheless, not completely arbitrary. On the one hand, they must be orthogonal to the hypercharge assignment in the sense of Eq.~\eqref{orthogonality}. Neglecting possible kinetic mixing effects, i.e. taking $f={\text{diag}}\,(g_a^{-2}\ldots \,g_d^{-2}\ldots)$, the orthogonality condition reads
\begin{equation}
\frac{a}{g_a^2}+\frac{3b}{g_c^2}-\frac{3d}{g_d^2} = 0 \,. 
\end{equation}
On the other hand, the vectors $\vec{v}^{\,(i)}$ must be properly normalized, yielding
\begin{equation}
\frac{g_Y^2}{36}\left(\frac{1}{g_a^2}+\frac{9}{g_c^2}+\frac{9}{g_d^2}\right)=1\,,
\end{equation}
\begin{equation}
\label{eq:norm}
\frac{a^2}{g_a^2}+\frac{b^2}{g_b^2}+\frac{c^2}{g_c^2}+\frac{d^2}{g_d^2}+\sum_{i=1}^m\frac{h_i^2}{g_{h_i}^2}=1\,,
\end{equation}
for the $Z$ and $Z'$ respectively. Notice that in the second expression the factor $\sum_{i=1}^m h_i^2/g_{h_i}^2$ encodes all the possible interactions of the $Z'$ with matter living in the hidden sector. Given the potential complexity of this sector, which we will not fully specify in this work, Eq.~\eqref{eq:norm} reduces to a bound on the visible sector couplings
\begin{equation}
\frac{a^2}{g_a^2}+\frac{b^2}{g_b^2}+\frac{c^2}{g_c^2}+\frac{d^2}{g_d^2}<1\,.
\label{eq:norm_visible}
\end{equation}

Furthermore, the couplings $g_i$ can be related to the SM gauge coupling constants by means of the following relations~\cite{Ibanez:2001nd,Ghilencea:2002da}:
\begin{equation}
g_a^2=\frac{g_3^2}{6}\,,\hspace{.3cm} g_b^2=\frac{g_2^2}{4}\,, \hspace{.3cm} \left(\frac{1}{g_a^2}+\frac{9}{g_c^2}+\frac{9}{g_d^2}\right)=36 g_Y^{-2}\,,
\end{equation}
where $g_3$ and $g_2$ refers to the $SU(3)_{QCD}$ and $SU(2)_L$ coupling constants, respectively.
These relations arise from the fact that $U(1)_A$ and $U(1)_B$ are just the center of the groups from
which the $SU(3)_{QCD}$ and $SU(2)_L$ gauge factors of the SM arise.\footnote{Note that these relations should be evaluated at the compactification scale. The running of the coupling constants from this scale to the electroweak scale, at which isospin violating properties of DM are defined, can be simply reabsorbed into the definition of the parameters $a$, $b$, $c$ and $d$.}

Now we have all the necessary information to build the couplings of the $Z'$ to the SM particles. 
In virtue of Eq.~\eqref{eq:charges} and table~\ref{tab:u1}, the left and right handed (first and second family) of quarks have the following couplings,
\begin{eqnarray}
\nonumber g_{u_L}^{Z'}=\left(a+b\right),\qquad  g_{u_R}^{Z'}=\left(-a+c\right)\,,\\ 
\,g_{d_L}^{Z'}=\left(a+b\right),\qquad g_{d_R}^{Z'}=\left(-a-c\right)\,,
\end{eqnarray}
which can be used to define the vectorial coupling as the sum of the left and right components,
\begin{eqnarray}
\nonumber C_{u}^V=g_{u_L}^{Z'} + g_{u_R}^{Z'}&=\left(b+c\right),\\
C_{d}^V=g_{d_L}^{Z'} + g_{d_R}^{Z'}&=\left(b-c\right),
\label{eq:cucd_vec}
\end{eqnarray}
and the axial coupling as the difference,
\begin{eqnarray}
\nonumber C^A_{u}=g_{u_L}^{Z'} - g_{u_R}^{Z'}&=\left(2a+b-c\right),\\
C^A_{d}=g_{d_L}^{Z'} - g_{d_R}^{Z'}&=\left(2a+b+c\right).
\label{eq:cucd_ax}
\end{eqnarray}
Similarly, according to Table~\ref{tab:u1}, for the third family of quarks the vectorial couplings are given by
\begin{eqnarray}
\nonumber C_{t}^V&=\left(-b-c\right),\\
C_{b}^V&=\left(-b+c\right)\,,
\label{eq:ctcb_vec}
\end{eqnarray}
whereas the axial couplings are given by
\begin{eqnarray}
\nonumber C^A_{t}&=\left(2a-b-c\right),\\
C^A_{b}&=\left(2a-b+c\right).
\label{eq:ctcb_ax}
\end{eqnarray}
Finally, for the three families of leptons the vector and axial couplings can be written as
\begin{eqnarray}
\nonumber C_{\ell}^V=\left(-b-c\right),\\
 C^A_{\ell}=\left(-b+c-2d\right),
\label{eq:cl_vec}
\end{eqnarray}
respectively. Note that in all cases, the vectorial couplings are independent of $a$ as well as of $d$, as was to be expected from the aforementioned interpretation of the charges $Q_A$ and $Q_D$ in terms of baryon and lepton number. The axial couplings, on the other hand do depend on $a$ and $d$. This fact will have a remarkable impact on the LHC bounds as we will see later.

\section{Isospin violation from the St\"uckelberg mechanism}\label{sec:isospinviolation}

As we have seen previously, the different charges of the SM particles under the $U(3)_A\times U(2)_B \times U(1)_C\times U(1)_D$ visible gauge group, together with the mixing of the corresponding abelian bosons, gave rise to very generic vector and axial couplings to the $Z'$ boson. As a consequence, a DM particle living in the hidden sector, $\psi$, will couple to each SM fermion through the $Z'$ in a different manner. This fact can be translated into a different coupling strength of $\psi$ to protons and neutrons, and thus, to a rather flexible amount of isospin violation $f_n/f_p$ ($a_n/a_p$). This is very important from the point of view of DM direct detection experiments~\cite{Feng:2011vu}.
 
Direct detection experiments are based on the elastic scattering of DM particles off nucleons inside an underground detector which shields it from cosmic rays. These experiments are tremendously sensitive to the recoil energy released by a nucleus of the target material when a DM particle hits it. Since the interaction between the nucleon and the DM particle occurs in the non relativistic limit (the relative velocity of the system in the lab frame is of the order of hundreds of km/s), the energy deposited in the detector after the collision is very small, of the order $\mathcal{O}(10)$~keV. Depending on the nature of the DM particles, and the mediator of its interaction with quarks, there exist many different operators that contribute to this interaction. For a Dirac fermion DM with a $Z'$ gauge boson mediator, its interactions with quarks can be divided into the so-called spin-independent (SI) interactions, arising from scalar and vector interactions with quarks, and spin-dependent (SD) interactions that originate from axial-vector interactions. Let us now analyse either cases separately.

\subsection{SI interactions}

The spin independent contribution to the total cross section of the DM-nucleus elastic scattering arises from scalar and vector couplings. For an interaction mediated by a vector boson exchange, the effective Lagrangian for the interaction of $\psi$ with nucleons (protons ($p$) and neutrons ($n$)) can be written as,
\begin{equation}
\mathcal{L}_{SI}^V=f_p(\bar{\psi}\gamma_\mu\psi)(\bar{p}\gamma^\mu p)+f_n(\bar{\psi}\gamma_\mu\psi)(\bar{n}\gamma^\mu n),
\end{equation}
where $f_p$ and $f_n$ are the vector couplings of $\psi$ to the protons and neutrons, respectively. These quantities depend on the nucleon quark content. For a vector interaction the only quarks that play a role are those of the valence (up and down), while for a scalar interaction the sea quarks are also important for the entire process.
Since the up and down quarks are not present in the proton and neutron in the same fraction, one can express $f_p$ and $f_n$ as follows~\cite{Chun:2010ve},
\begin{equation}
f_p = 2b_u + b_d, \quad f_n=b_u + 2b_d,
\label{eq:fnfp}
\end{equation}
where $b_u$ and $b_d$ are the effective vector couplings of the up and down quarks to the DM particles.\footnote{ Not to be confused with the coupling $b$ associated with the $U(1)_B$ symmetry.} After integrating out the $Z'$ boson, these couplings can be easily written as,
\begin{equation}
b_{(u,d)}=\frac{hC^V_{(u,d)}}{2m_{Z'}^2}\,,
\end{equation}
with $h$ being the coupling strength of the $Z'$ boson to $\psi$, and $m_{Z'}$ the mass of the lightest $Z'$ boson.

\begin{figure}[tpb]
      \begin{center}
\scalebox{0.9}{
            \hspace{-0.65cm}\includegraphics[scale=0.45]{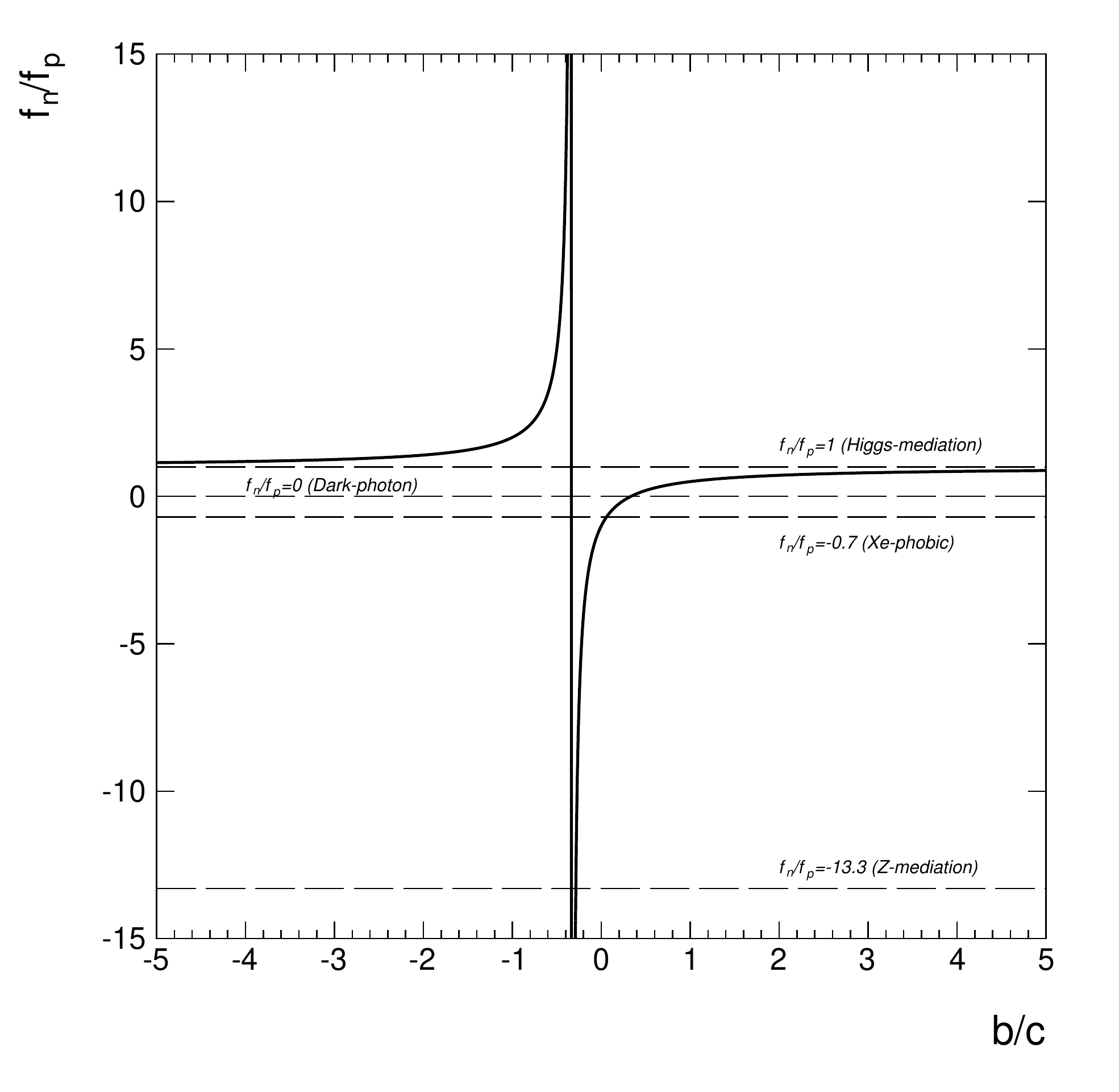}
            \includegraphics[scale=0.45]{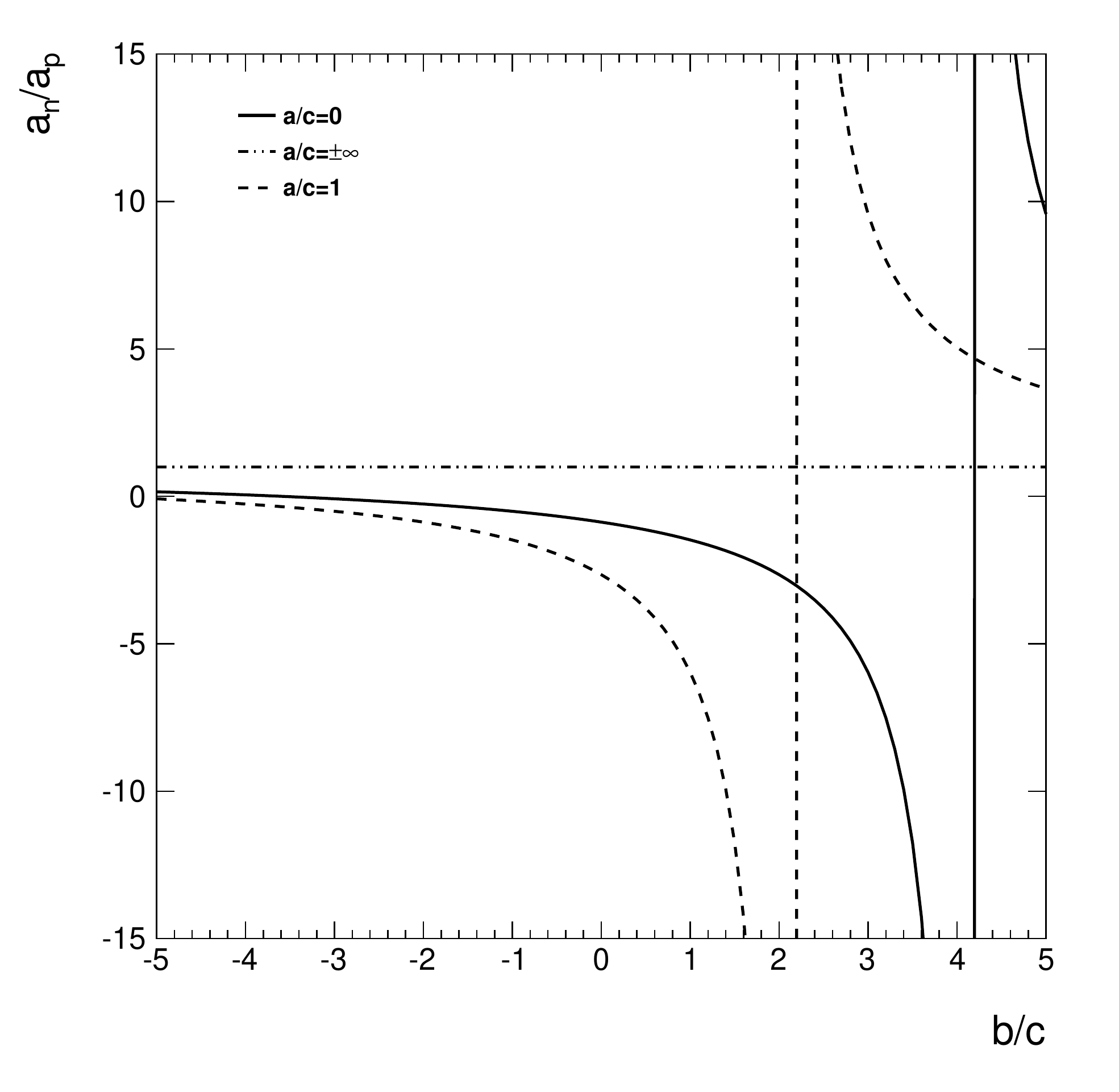}
              }
\caption{\small Left: Amount of isospin violation for SI interactions, $f_n/f_p$, as a function of $b/c$ (solid line). Some representative values of $f_n/f_p$ are shown as horizontal dashed lines. Right: Ratio between the coupling of DM to neutrons and protons, $a_n/a_p$, for the SD interactions as a function of $b/c$. For $a/c$ we have taken different limits, $a/c\rightarrow 0$ (solid line),  $a/c\rightarrow\pm\infty$ (dot dashed line), and $a/c=1$ (dashed line).}
      \label{fig:fnfpbc}
      \end{center}
\end{figure}

Using now the expressions for the vector couplings of the $Z'$ to up and down quarks, given in Eq.~\eqref{eq:cucd_vec}, it is straightforward to deduce that,
\begin{eqnarray}
\nonumber b_u=\frac{h C^V_u}{2m^2_{Z'}}=\frac{h}{2m^2_{Z'}}\left(b+c\right),\\ b_d=\frac{h C^V_d}{2m^2_{Z'}}=\frac{h}{2m^2_{Z'}}\left(b-c\right).
\end{eqnarray}
These two expressions make obvious that in this framework the ratio between the coupling of $\psi$ to protons and neutrons i.e. the amount of isospin violation $f_n/f_p$, according to Eq.~\eqref{eq:fnfp}, is given by
\begin{equation}
\frac{f_n}{f_p}=\frac{\left(3b-c\right)}{\left(3b+c\right)}=\frac{\left(3b/c-1\right)}{\left(3b/c+1\right)}\,.
\label{eq:fnfp_ratio}
\end{equation}
Interestingly, the the total amount of isospin violation depends exclusively on the ratio between the parameters $b$ and $c$ which, as mentioned before, are continuous and different from zero, generating a ratio $f_n/f_p$ different from $\pm1$. This is a consequence of the introduction of $U(1)$ gauge groups in the visible sector to reproduce the global symmetries of the SM. In particular, the parameter $b$ corresponds to a chiral $U(1)$ symmetry of the Peccei-Quinn type, with mixed $SU(3)$ anomalies; while $c$ is related precisely to the weak isospin symmetry $U(1)_C$~\cite{Ibanez:2001nd}. Isospin violation and chirality are the key properties why these new groups generate a general isospin violation in the currents related to the $Z'$ interaction.

In Figure~\ref{fig:fnfpbc} (left panel) the quantity $f_n/f_p$ is shown as a function of $b/c$ according to Eq.~\eqref{eq:fnfp_ratio}. We have shown some noteworthy theoretical benchmark values of this ratio as well, like $Z$ mediation and dark photon scenarios, $f_n/f_p=-13.3$ and $f_n/f_p=0$, respectively. The value of $f_n/f_p\approx-0.7$ is the so-called Xe-phobic dark matter scenario to which Xe-based detectors are poorly sensitive\footnote{This is a consequence of the ratio between the number of protons and neutrons in xenon isotopes}. Interestingly, we notice that our construction naturally generates isospin violating couplings $f_n/f_p\neq 1$ for any value of the parameters $b$ and $c$. These parameters are expected to be of the same order, $|b/c|\sim{\cal O}(1)$, which defines a region in which the value of $f_n/f_p$ is subject to important changes (for values around $b/c=-1/3$). This precisely highlights the flexibility in the isospin violation patterns found in these constructions. 

All this together can be taken as a clear and testable prediction of this kind of constructions. It also would be distinguishable from other hidden DM scenarios. For instance, if the portal between the visible and the hidden sector occurs via a Higgs boson, the value of $f_n/f_p$ would be generally $1$, since the Higgs boson can not differentiate chiralities of the quarks.\footnote{ In type II 2HDM for $\tan\beta\approx1$ there can be deviations~\cite{Drozd:2014yla}.}

It is worth noting that, although the type of constructions we are considering, based on the visible gauge group $U(3)_A\times U(2)_B \times U(1)_C\times U(1)_D$, lead to a flexible amount of isospin violation (generically $f_n/f_p\neq \pm1$),\footnote{In the models we discuss, the values $f_n/f_p=\pm 1$ can only be reached in the limits $b/c\rightarrow 0$ and $b/c\rightarrow\infty$, which although not excluded, are not particularly preferred. This provides a remarkable and potentially measurable distinction of these constructions from other portals.} there is a well known class of alternative type II string models in which the gauge group $U(2)_B$ is replaced by $USp(2)_B\cong SU(2)_B$~\cite{Cremades:2003qj}. In such models, the $U(1)_B$ factor, which was crucial in our discussion, is absent. One could realise the St\"uckelberg portal scenario in such constructions, and follow steps similar as the ones we have taken here. The only difference one would find is that the parameter $b$ would  be identically zero, and hence that the DM interactions with the nucleons would automatically satisfy $f_n/f_p\equiv -1$.

\subsection{SD interactions}

Let us now move to consider the case of SD interactions. As we have mentioned above, these interactions arise from the axial-vector couplings of DM to protons and neutrons, and thus, occur when the DM particles have a spin different from zero. In terms of the effective Lagrangian we can write,
\begin{equation}
\mathcal{L}_{SD}=a_p(\bar{\psi}\gamma_\mu \gamma_5\psi)(\bar{p}\gamma^\mu \gamma_5 p)+a_n(\bar{\psi}\gamma_\mu \gamma_5\psi)(\bar{n}\gamma^\mu \gamma_5 n),
\end{equation}
where the parameters $a_{p(n)}$ are the couplings of DM to protons (neutrons), and can be expressed in the following way~\cite{Cerdeno:2010jj},
\begin{equation}
a_p=\sum_{q=u,d,s}\frac{\alpha_q^A}{\sqrt{2}G_F}\Delta_q^p=\frac{h}{2\sqrt{2}G_F m_{Z'}^2}\left[C^A_u\Delta_u^p + C_d^A(\Delta_d^p + \Delta_s^p) \right],
\end{equation}
\begin{equation}
a_n=\sum_{q=u,d,s}\frac{\alpha_q^A}{\sqrt{2}G_F}\Delta_q^n=\frac{h}{2\sqrt{2}G_F m_{Z'}^2}\left[ C_u^A\Delta_u^n +  C_d^A(\Delta_d^n + \Delta_s^n) \right]\,,
\end{equation}
where $\alpha_q^A$ is the effective axial coupling of DM to quarks and $G_F$ denotes the Fermi coupling constant. Operators for axial-vector
interactions in the nucleon are related to those involving quarks through the quantities $\Delta_q^{p(n)}$, which relate the spin of the nucleon to the operator $\langle p(n)|\bar{q}\gamma^\mu \gamma_5 q|p(n)\rangle$. For these we have taken the values from Ref.~\cite{Belanger:2013oya}.

Now, we can take the ratio between the coupling to protons and neutrons, which gives
\begin{equation}
\frac{a_n}{a_p}=\frac{\Delta_u^n + \frac{2a/c+b/c+1}{2a/c+b/c-1}(\Delta_d^n+\Delta_s^n)}{\Delta_u^p + \frac{2a/c+b/c+1}{2a/c+b/c-1}(\Delta_d^p+\Delta_s^p)}.
\label{eq:anap_ratio}
\end{equation}
As one can see from the previous expression, unlike for $f_n/f_p$, this ratio also depends on $a/c$ not only on $b/c$, and hence, there is one more degree of freedom respect to the SI case. 

\begin{figure}[tpb]
\centering
\scalebox{0.9}{
            \includegraphics[scale=0.5]{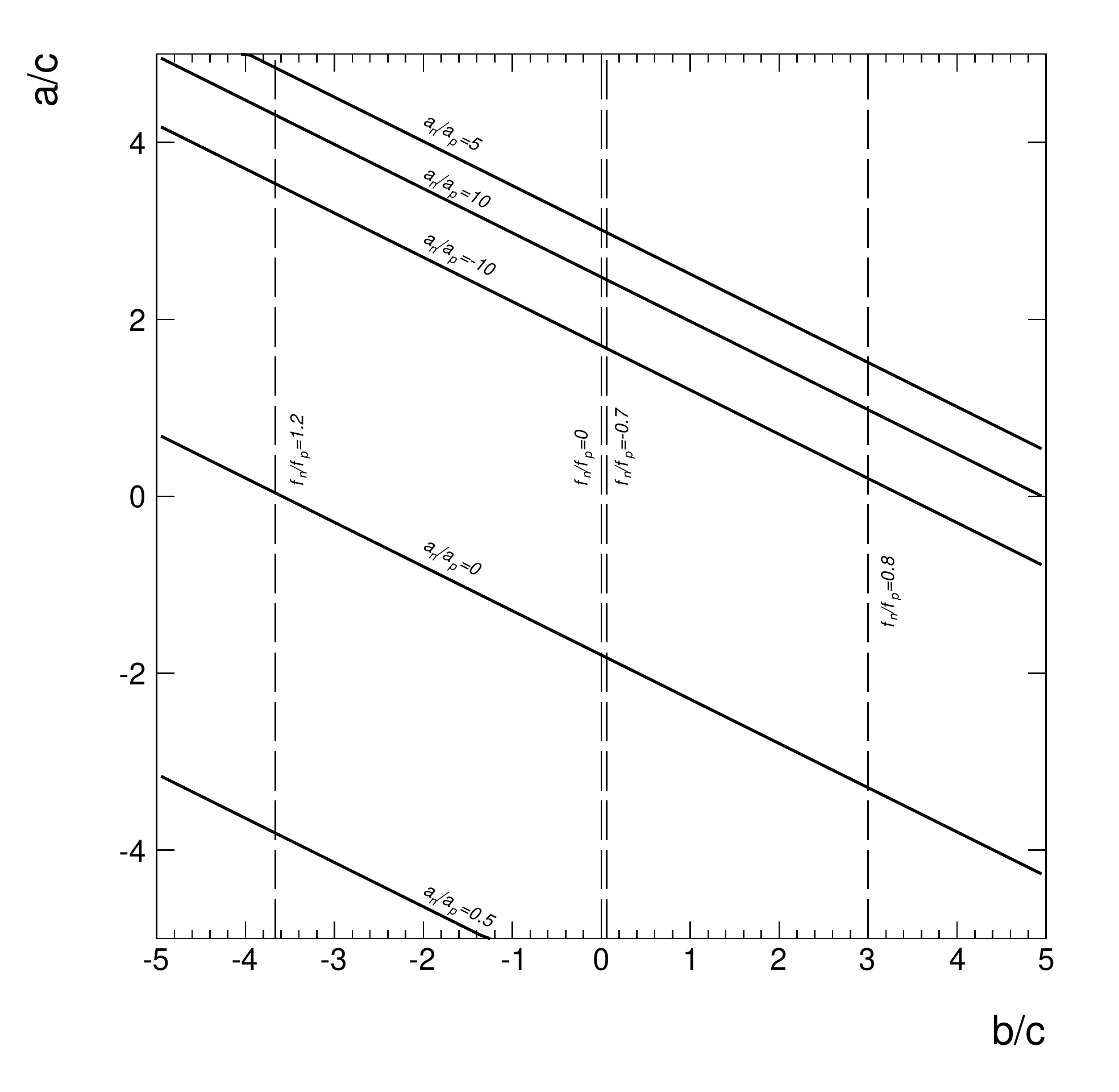}
               }\caption{\small Ratio between the coupling of DM to neutrons and protons for the SD interactions as a function of $b/c$. For $a/c$ we have take different limits, $a/c\rightarrow 0$ (black line),  $a/c\rightarrow \infty$ (blue line), and $a/c=1$ (gray line).}\label{fig:contours}
\end{figure}

In Figure~\ref{fig:fnfpbc} (right panel), the ratio $a_n/a_p$ is depicted as a function of $b/c$ according to Eq.~\eqref{eq:anap_ratio} for different values of the ratio $a/c$. In the limit of $a/c\rightarrow\infty$ (dot dashed line), we find the case of $a_n/a_p=1$, similar to the case of the SI interactions in the limit $b/c\rightarrow\infty$. While, for the cases $a/c\rightarrow 0$ (solid line) and $a/c= 1$ (dashed line), the values of $a_n/a_p$ are generally different from $\pm1$. Notice that in this case one can also define the Xe-phobic scenario for $a_n/a_p$. However, it depends on the ratio between the zero momentum expectation values of the spin for protons and neutrons in xenon which are of the order of $\mathcal{O}(10^{-2})$ (using the latest calculations~\cite{Klos:2013rwa}), and for simplicity it is not included in Figure~\ref{fig:fnfpbc}. 

Finally, in order to rearrange the results for both SI and SD interactions, in Figure~\ref{fig:contours} we show the plane $a/c$ versus $b/c$. As we have seen before, these two ratios determine the amount of isospin violation in DM interactions for both the SI and SD contributions. On the one hand, the dashed vertical lines represent some values of $f_n/f_p$, which are independent of $a/c$, as in  Figure~\ref{fig:fnfpbc}. On the other hand, the solid lines denote some values for $a_n/a_p$. Remarkably, in the region shown, where the values of $a$, $b$ and $c$ are in general of the same order, the DM interactions are isospin violating in both types of interactions. Furthermore, we see that very high values of the neutron component (with respect to  the proton component) can be reached, although, the variation of either $f_n/f_p$ or $a_n/a_p$ is very abrupt in this region (see also Figure~\ref{fig:fnfpbc}). This is important for direct detection experiments that use target materials in which the ratio between the neutron and proton contribution is significantly different than one. For instance, in Xe-based detectors such as LUX, the SD component is dominated by the neutron scatterings due to the dominance of the neutrons in the total spin of the $^{129}$Xe and $^{131}$Xe isotopes.

\section{Isospin violating DM in light of the LHC and LUX results}\label{sec:bounds}

As we have shown previously, this kind of constructions generally predict isospin violating DM. The relations between the proton and neutron contributions for SI and SD interactions depend on the couplings $a$, $b$ and $c$, and more specifically, in their relations. 

According to Eqs.~\eqref{eq:cucd_vec}-\eqref{eq:cl_vec}, all couplings of the $Z'$ to SM fermions can be written in terms of the four parameters $a\,,b\,,c\,,d$. In light of this, it is obvious that certain combinations of these parameters will affect the predicted values of some constrained experimental observables. Furthermore, as pointed out in Section~\ref{sec:SMcouplings}, there are some constraints on these parameters that come from the building of the $Z$ and $Z'$ bosons in this model. This section is aimed at exploring the impact that these constraints have on the allowed values of $a/c$ and $b/c$, and hence, on the experimentally allowed values of $f_n/f_p$ and $a_n/a_p$. Needless to say, these regions will depend on certain assumptions on the DM mass and its coupling $h$, the $Z'$ mass, $d/c$, and $c$, and for this reason we will concentrate on six representative benchmark (BM) points. The values used for each of these parameters are shown in Table~\ref{tab:BM}. 

It is legitimate to ask whether the mass scales that appear in such BMs can arise in consistent string compactifications. As we have already mentioned at the end of section~\ref{subsec:stringtheory}, $Z'$ masses of the order of the TeV, although not generic, can be achieved in several ways without much difficulty. On the other hand, notice that the DM particles are charged, often chirally, not only under $U(1)$ hidden groups, but also under non-abelian factors, i.e. the $G_{\rm h}$ in Eq.~\eqref{stuckportal}. Therefore, the mass of the field $\psi$ is related to possible strong coupling dynamics and symmetry breaking patterns (e.g. a hidden Higgs mechanism) of the hidden non-abelian gauge sector. In this sense, it is quite natural to consider DM masses in the GeV-TeV range, at least as natural as having visible sectors reproducing the masses of the SM particles.

\begin{table}[t]
  \centering
  \begin{tabular}{|c|c|c|c|c|c|}
    \hline
     & $c$ & $d/c$ & $h$ & $m_{\psi}$ (GeV) & $m_{Z'}$ (TeV) \\[3pt]
    \hline
    BM1 & $0.01$ & $1$ & $0.1$ & 50 & 1 \\[3pt]
        \hline
    BM1a & $0.01$ & $2$ & $0.05$ & 50 & 1 \\[3pt]
        \hline
    BM2 & $0.1$ & $3$ & $0.5$  & 500 & 3 \\[3pt]
            \hline
    BM2a & $0.05$ & $5$ & $0.25$ & 500 & 3 \\[3pt]
        \hline
    BM3 &$0.1$ & $1$ & $0.1$ & 2000 & 3 \\[3pt]
            \hline
    BM3a & $0.25$ & $1$ & $0.2$ & 2000 & 3 \\[3pt]
    \hline
    \hline
  \end{tabular}
  \caption{Input parameters for each BM point.}
  \label{tab:BM}
\end{table}

Given the potential complexity of the matter and gauge structure of the hidden sector, it seems reasonable to asume that there could be some mechanisms, either thermal or non-thermal, to account for the relic abundance of $\psi$ other than annihilation through the $Z'$ channel. This highlights the dependence of the DM abundance on the particular details of the hidden sector dynamics, which we want to keep as generic as possible. Nevertheless, it is worth mentioning that, in general, annihilation cross sections through the $Z'$ channel are lower than the thermal value, and thus  indirect detection bounds on the annihilation cross section are generally far from our predictions.

The only contributions to the phenomenology of the model that do not depend on any further assumption on the hidden sector are direct DM searches and LHC searches for resonances\footnote{ Hadronic decays~\cite{Fayet:2007ua,Frandsen:2011cg}  and the muon anomalous magnetic moment~\cite{Chun:2010ve,Teubner:2010ah} do not depend either on any further assumption and can affect the allowed values of parameters $a$, $b$ and $c$. However, we have checked that these constraints are not competitive with LUX and LHC in the region of the parameter space considered in this work.}. The former only depends on the coupling of $\psi$ to quarks by the exchange of a $Z'$ boson (see Figure~\ref{drawing}), while the latter depends on the coupling of $Z'$ to SM particles (quarks and leptons) and the coupling $h$. In the following, we will determine the experimentally allowed regions of $a/c$ and $b/c$ in the six BM points shown in Table~\ref{tab:BM} taking into account the limits from LUX and the LHC.

\subsection{LUX and LHC limits}

The recent null results of the LUX collaboration~\cite{Akerib:2013tjd} have placed a very stringent upper limit on the elastic scattering of DM off protons, reducing significantly the parameter space allowed in many theories that provide DM candidates. This limit has been extracted by assuming a scalar DM candidate (zero SD contribution) and $f_n/f_p=1$, which are the typical assumptions that the collaborations use in order to compare their results within a unified framework. However, this prevents us from using this result directly, since none of these two assumptions hold for the DM candidate analysed in this work. Therefore, in order to implement this bound properly we have simulated the LUX experiment, and we have calculated for a given point of the parameter space if it is allowed at 90\% C.L. using the Yellin's maximum gap method~\cite{Yellin:2002xd}. To such end, we calculate the predicted total number of events in LUX considering the SI and SD components and computing the $f_n/f_p$, $a_n/a_p$ ratios. To calculate the 90\% C.L. exclusion using the maximum gap method, we consider that LUX experiment has observed zero candidate events in the signal region\footnote{ Actually, LUX observed one candidate event that was marginally close to the background region in the $\log_{10}(S2/S1)-S1$ plane. Thus our result of the exclusion is closer to the actual LUX limit when considering zero observed events.}.

To calculate the total number of expected signal events in a Xe-based detector we have followed the prescription of Ref.~\cite{Aprile:2011hx} in the S1 range 2-30 PE for an exposure of 10065 kg days, using the acceptance shown in the bottom of Fig. 1 of Ref.~\cite{Akerib:2013tjd} plus an extra $1/2$ factor to account for the 50\% of nuclear recoil acceptance. We use the S1 single PE resolution to be $\sigma_{PMT} = 0.37$
PE~\cite{pmt}, a 14\% of photon detection efficiency, and the absolute scintillation efficiency digitized from Ref.~\cite{Akerib:2013tjd}. For the DM speed distribution, we use the standard isothermal Maxwellian velocity distribution, with $v_0 = 220$~km/s, $v_{esc} = 544$~km/s, $\rho_0 = 0.3$~GeV/cm$^3$
and $v_e = 245$~km/s, as the one used by the LUX collaboration~\cite{Akerib:2013tjd}. As pointed out in Ref.~\cite{Cerdeno:2012ix} the effect of the form factors can also induce important differences in the expected number of events. In this work we use the Helm factor for the SI component and the SD structure functions given in Ref.~\cite{Klos:2013rwa} for the SD component.

To show explicitly the dependence of the SI and SD elastic scattering cross sections on the parameters of the model, namely, on the ratios $a/c$ and $b/c$, let us write them as,
\begin{equation}
\sigma_p^{SI}=\frac{4}{\pi}\mu_p^2f_p^2=\frac{\mu_p^2h^2}{\pi m^4_{Z'}}\left(3b+c\right)^2,
\label{eq:si}
\end{equation}
\begin{equation}
\sigma_p^{SD}=\frac{24G_F^2}{\pi}\mu_p^2a_p^2=\frac{3\mu_p^2h^2}{\pi m_{Z'}^4}[(2a+b-c)\Delta_u^p+(2a+b+c)(\Delta_d^p+\Delta_s^p)]^2.
\label{eq:sd}
\end{equation}
Notice that in order to calculate the neutron contributions one has to multiply by $(f_n/f_p)^2$ the SI component and by $(a_n/a_p)^2$ the SD component, whose expressions are given in Eqs.~\eqref{eq:fnfp_ratio} and \eqref{eq:anap_ratio}. Let us mention at this point the existing relation between the SI and the SD elastic cross sections. From the previous equations, and the corresponding neutron counterparts, one can easily see that the contribution from the SD cross section to the total number of expected events dominates if $|a/c|\gg |b/c|$ and $|a/c|\gg1$. However, for a given of $c$ the ratio $a/c$ cannot be arbitrarily large due to the normalization Eq.~\eqref{eq:norm_visible}. In fact, it can be shown that for the SD component to be dominant in LUX for the range of $m_{\psi}$ considered and when $|b/c|<5$ (the region shown in the figures) then $a/c\gtrsim100$. Using the values of $c$ shown in Table~\ref{tab:BM}, such high values of $a/c$ do not satisfy the Eq.~\eqref{eq:norm_visible}, and hence, they are not considered.

The production and the subsequent decay of a $Z'$ boson into SM particles might leave distinctive signal of new physics that can be searched at colliders, and in particular at the LHC. The ATLAS detector at the LHC searched for high mass resonances decaying into a $\mu^+\mu^-$ or an $e^+e^-$ pair for energies above the $Z$ pole mass, at a center of mass energy $\sqrt{s}=8$~TeV and luminosities of $20.5$~fb$^{-1}$ and $20.3$~fb$^{-1}$ for dimuons and dielectrons resonances, respectively~\cite{Aad:2014cka}. These results are consistent with the SM predictions allowing to place an upper limit on the signal cross section times the corresponding branching fraction of the process $pp\rightarrow Z'\rightarrow\mu^+\mu^-(e^+e^-)$.\footnote{Although these results can be used to place constraints on other models of new physics, we are interested in its application for the search of a $Z'$ boson.} There are also searches for dijet resonances and monojets plus missing energy that receive additional contributions from the presence of a $Z'$ boson, both at the LHC and Tevatron colliders, and hence, they can be used to place constraints on this kind of models as well~\cite{Aaltonen:2008dn,Aad:2011fq,CMS-dijet}.

In the model presented here, the coupling of the $Z'$ boson to leptons and quarks contributes to the appearance of dimuon, dielectron and dijet resonances, and thus, these searches can constraint the parameter space. In order to include these bounds to determine which regions are allowed in light of these searches, we have followed the approach given in Ref.~\cite{Carena:2004xs}. In the narrow width approximation, the dilepton production in proton-proton collisions mediated by the $Z'$ can be written as,
\begin{equation}
\sigma_{l^+l^-}\simeq\left(\frac{1}{3}\sum_{q}\frac{dL_{q\bar{q}}}{dm_{Z'}^2}\times\widehat{\sigma}(q\bar{q}\rightarrow Z')\right)\times {\rm BR}(Z'\rightarrow l^+l^-)\,,
\end{equation}
where $dL_{q\bar{q}}/dm_{Z'}^2$ denotes the parton luminosities, $\widehat{\sigma}(q\bar{q}\rightarrow Z')$ is the peak cross section for the $Z'$ boson, and ${\rm BR}(Z'\rightarrow l^+l^-)$ is the branching ratio for the $Z'$ decaying into a lepton pair. A close inspection of the previous expression reveals that there is a part which only depends on the model parameters, and the remaining part that only depends on the kinematics of the process. Hence, it can be factorized as,
\begin{equation}
\sigma_{l^+l^-}=\frac{\pi}{48s}\mathcal{W}_{Z'}\left( s,m_{Z'}^2\right)\times {\rm BR}(Z'\rightarrow l^+l^-)\,,
\label{eq:sigma1}
\end{equation}
where the function $\mathcal{W}_{Z'}$ is given by:
\begin{equation}
\mathcal{W}_{Z'}=\sum_{q=u,d,c,s}c_{q}\omega_{q}\left( s,m_{Z'}^2\right)\,.
\end{equation}
The coefficients $c_q$ are the sums of the squares of the vector and axial couplings, $(C^V_q)^2+(C^A_q)^2$, to the corresponding quarks. Notice that we do not include the contributions from the bottom and top quarks, since they can be safely neglected in the production process. In this limit, provided that the first and second quark families share the same charges under the $U(3)_A\times U(2)_B \times U(1)_C\times U(1)_D$ gauge symmetry group (see Section~\ref{sec:SMcouplings}), the function $\mathcal{W}_{Z'}$ can be written as a sum of the up and down doublet components of the quarks as
\begin{equation}
\mathcal{W}_{Z'}=c_{up}\omega_{up}\left( s,m_{Z'}^2\right)+c_{down}\omega_{down}\left( s,m_{Z'}^2\right)\,.
\label{eq:WZ}
\end{equation}
In the previous expression we have reabsorbed a factor 2 in the definition of the $\omega$ functions. This factor corresponds to the sum of the up and charm quarks contribution to the up component and, in the same way, for the down and strange quarks for the down component.

Using the equations~\eqref{eq:sigma1} and \eqref{eq:WZ} one can easily write the production cross section of a dilepton pair mediated by the $Z'$ in proton proton collisions at leading order (LO) as,
\begin{equation}
\sigma^{LO}_{l^+l^-}=\left[c_{up}\tilde{\omega}_{up}\left( s,m_{Z'}^2\right)+c_{down}\tilde{\omega}_{down}\left( s,m_{Z'}^2\right)\right] \times {\rm BR}(Z'\rightarrow l^+l^-)\,,
\end{equation}
where $\tilde{\omega}_{up,down}=(\pi/48s) \omega_{up,down}$. To extract the functions $\tilde{\omega}$ at $\sqrt{s}=8$~TeV we have benefited from {\tt CalcHEP 3.6.22}~\cite{Belyaev:2012qa} using the parton distribution functions {\tt CTEQ6L} to be consistent with the LHC analysis~\cite{Aad:2014cka}. Furthermore, in order to include Next-to-LO effects, we have used the K-factor given in Ref.~\cite{Accomando:2010fz}. Remarkably, this approach can be used to calculate not just the bounds for dilepton but also for dijet resonances just by substituting ${\rm BR}(Z'\rightarrow l^+l^-)$ (valid for dilepton in the previous expressions) by ${\rm BR}(Z'\rightarrow q\bar{q})$, where a sum over all quarks (except for the top quark~\cite{CMS-dijet}) must be performed. Finally, to include properly the dijet resonance searches, the cross section times the branching fraction must be multiplied by a factor $A=0.6$ which accounts for the efficiency of the detector~\cite{CMS-dijet}.

Before moving to the phenomenological analysis of the BM points, let us write the partial widths of the $Z'$ boson decay into SM particles and $\psi$ as a function of the model parameters as,
{\small{\begin{eqnarray}
	\Gamma_{l\bar{l}} &=& \frac{m_{Z'}}{12\pi}c^2\left[\left(1+\frac{b}{c}\right)^2\left(1+\frac{2m_l^2}{m_{Z'}^2}\right)+\left(1-\frac{b}{c}-2\frac{d}{c}\right)^2\left(1-\frac{4m_l^2}{m_{Z'}^2}\right)\right]\sqrt{1-\frac{4m_l^2}{m_{Z'}^2}},\\
		\Gamma_{\nu\bar{\nu}} &=& \frac{m_{Z'}}{6\pi}c^2\left(\frac{b}{c}+\frac{d}{c}\right)^2,\\
	\Gamma_{u\bar{u}(c\bar{c})} &=& \frac{m_{Z'}}{4\pi}c^2\left[\left(1+\frac{b}{c}\right)^2\left(1+\frac{2m_{u(c)}^2}{m_{Z'}^2}\right)+\left(1+\frac{b}{c}+2\frac{a}{c}\right)^2\left(1-\frac{4m_{u(c)}^2}{m_{Z'}^2}\right)\right]\sqrt{1-\frac{4m_{u(c)}^2}{m_{Z'}^2}},\\
	\label{eq:gammauu}
		\Gamma_{d\bar{d}(s\bar{s})} &=& \frac{m_{Z'}}{4\pi}c^2\left[\left(1-\frac{b}{c}\right)^2\left(1+\frac{2m_{d(s)}^2}{m_{Z'}^2}\right)+\left(1+\frac{b}{c}+2\frac{a}{c}\right)^2\left(1-\frac{4m_{d(s)}^2}{m_{Z'}^2}\right)\right]\sqrt{1-\frac{4m_{d(s)}^2}{m_{Z'}^2}},\\
	\label{eq:gammadd}
		\Gamma_{t\bar{t}} &=& \frac{m_{Z'}}{4\pi}c^2\left[\left(1-\frac{b}{c}\right)^2\left(1+\frac{2m_{t}^2}{m_{Z'}^2}\right)+\left(1+\frac{b}{c}-2\frac{a}{c}\right)^2\left(1-\frac{4m_{t}^2}{m_{Z'}^2}\right)\right]\sqrt{1-\frac{4m_{t}^2}{m_{Z'}^2}},\\
		\Gamma_{b\bar{b}} &=& \frac{m_{Z'}}{4\pi}c^2\left[\left(1+\frac{b}{c}\right)^2\left(1+\frac{2m_{b}^2}{m_{Z'}^2}\right)+\left(1-\frac{b}{c}+2\frac{a}{c}\right)^2\left(1-\frac{4m_{b}^2}{m_{Z'}^2}\right)\right]\sqrt{1-\frac{4m_{b}^2}{m_{Z'}^2}},\\
	    \Gamma_{\psi\bar{\psi}} &=& \frac{m_{Z'}}{6\pi}h^2\left(1-\frac{m_{\psi}^2}{m_{Z'}^2}\right)\sqrt{1-\frac{4m_{\psi}^2}{m_{Z'}^2}},
\end{eqnarray}}}
where $l$ and $\nu$ refer to the three families of leptons and neutrinos, respectively. These expressions and the SM couplings of the $Z'$, given in Section~\ref{sec:SMcouplings}, allow us to evaluate the LHC bounds as a function of the parameters $a/c$ and $b/c$.

\subsection{Results}

\begin{figure}[tpb]
\centering
\scalebox{0.9}{
            \includegraphics[scale=0.44]{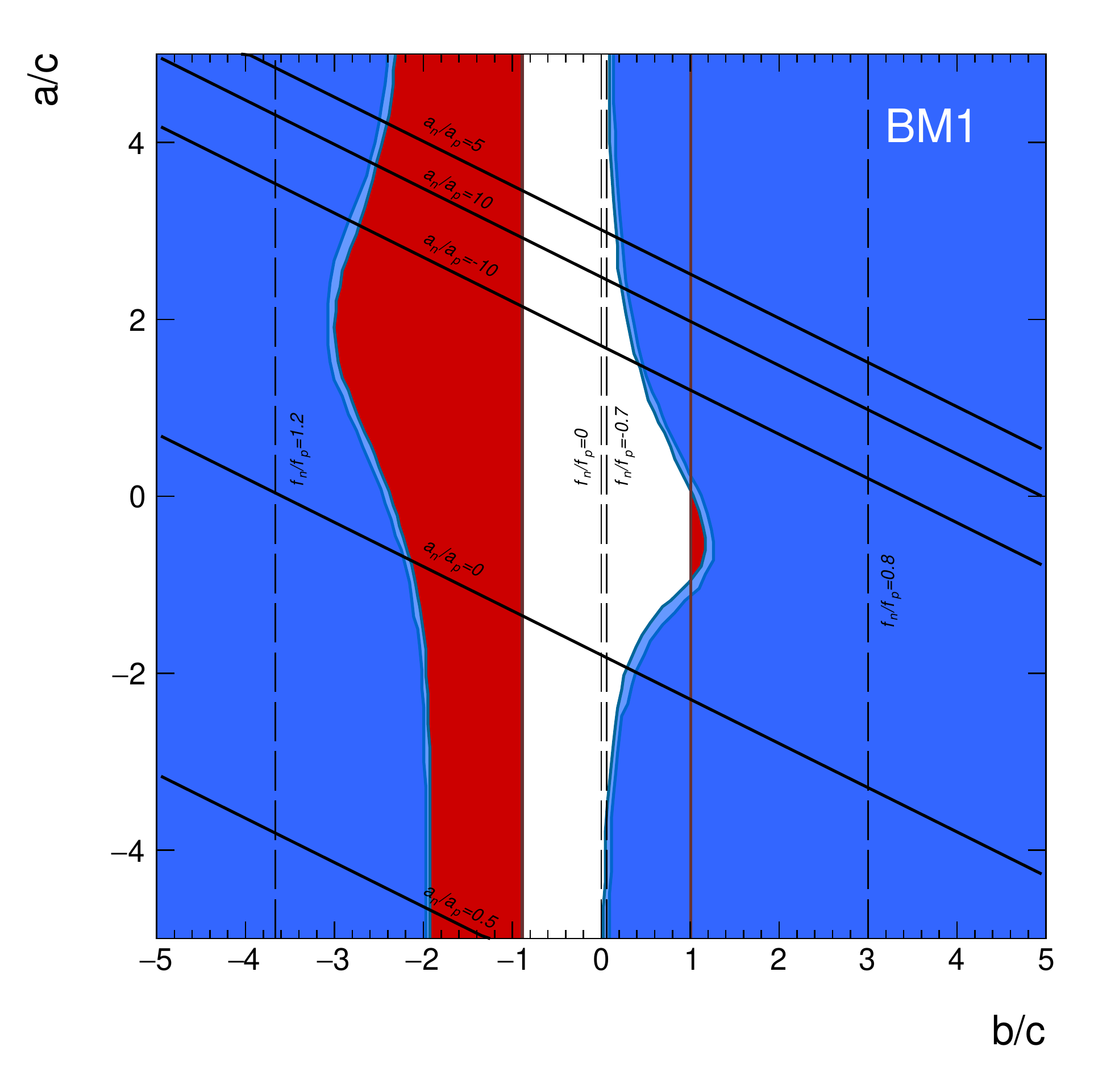}
                        \includegraphics[scale=0.44]{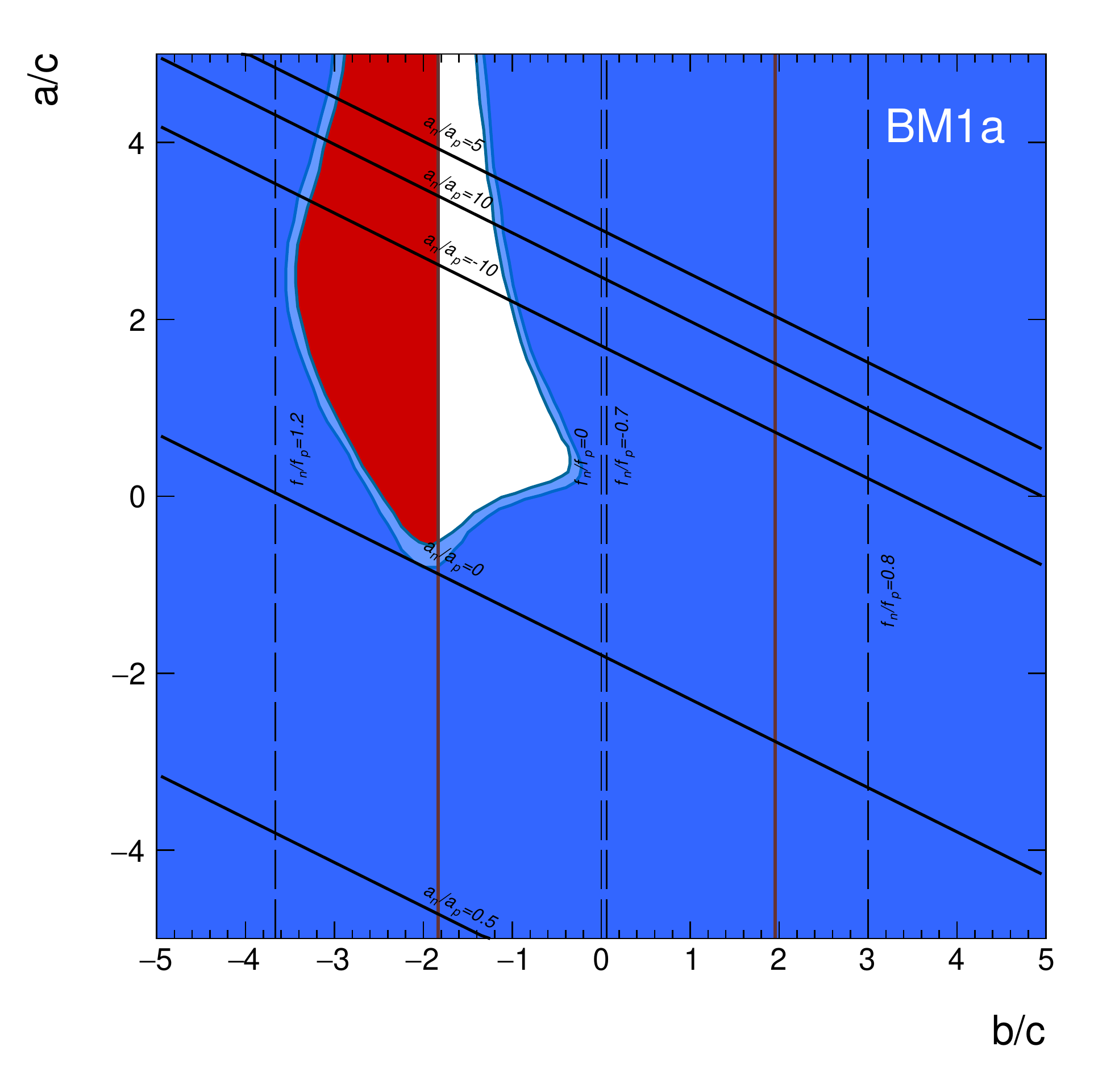}
               }\caption{\small $a/c$ versus $b/c$ for BM1 (left) and BM1a (right). As in figure~\ref{fig:contours} we show some values of the ratios $f_n/f_p$ and $a_n/a_p$. The LUX bound excludes all the region depicted in red. LHC bounds rule out different regions for $e^+e^-$ (light blue) and $\mu^+\mu^-$ resonances (darker blue). In this BM point, dijet resonances do not constrain.}\label{fig:BM1}
\end{figure}

Let us start analysing BM1 and BM1a. These BM points correspond to a low mass dark matter candidate, with a mass of 50 GeV and a $Z'$ boson of 1 TeV. In figure~\ref{fig:BM1} we show the plane $a/c$-$b/c$ with some values of the ratios $f_n/f_p$ and $a_n/a_p$ for BM1 (left panel) and BM1a (right panel). We have superimposed the 90\% C.L. LUX exclusion region (shown in red) that rules out high values of $|b/c|$, while in blue we show the exclusion regions from the LHC searches for $e^+e^-$ (light blue) and $\mu^+\mu^-$ resonances (darker blue). As we have anticipated previously, the LUX limit does not depend on the specific value of $a/c$ since in this region of the parameter space the SI contribution of the elastic scattering dominates over the SD one. For BM1, LUX excludes the regions $b/c\lesssim -0.9$ and $b/c\gtrsim 1.0$, which correspond to the regions in which the proton and neutron components of the SI elastic cross section are similar, $|f_n/f_p|\approx 1$. To understand this behaviour note that the proton contribution given in Eq.~\eqref{eq:si} decreases very fast around $b/c=-1/3$, faster than $f_n/f_p$ (due to $f_p^2$). This means that, although in the allowed region the neutron contribution to the SI cross section dominates with values of $f_n/f_p$ that can be very large (see also left panel of Figure~\ref{fig:fnfpbc}), it also decreases, and thus, the LUX limit weakens. For BM1a, since the value of $h$ has been decreased respect to BM1, the coupling of $\psi$ to the $Z'$ also diminishes and then the LUX limits are able to constrain much less parameter space, namely, it rules out the region $|b/c|\gtrsim1.9$.

Unlike direct detection limits, LHC bounds depend on the value of $a/c$. First of all, we show that for BM1 when $|a/c|\lesssim2$, both $e^+e^-$ and $\mu^+\mu^-$ bounds are less stringent. This can be understood from Eqs.~\eqref{eq:gammauu} and \eqref{eq:gammadd}. The second term in both expressions is minimized when $b/c\approx-2a/c$ which is translated into a minimization of the production cross section of the $Z'$ (see also Eq.~\eqref{eq:WZ}) and thus, both limits are less stringent. Besides, LHC limits are stronger for positive values of $b/c$ as a consequence of the dominance of the $\tilde{\omega}_{up}$ function over the corresponding function of the down component and hence, the production through the up component cancels out the first term of Eq.\eqref{eq:gammauu} for $b/c\approx-1$. LHC limits are stronger for BM1a for two reasons. First, the increasing of $d/c$ makes the $Z'$ coupling to leptons higher and then the corresponding branching ratio is increased. Second, a smaller value of $h$ makes the $Z'$ boson less \textit{invisible}, which is translated into an increase of both, its production cross section and its branching ratio into SM particles. Remarkably, LHC limits rule out a big portion of the parameter space allowed by LUX, including the Xe-phobic value of $f_n/f_p$, and it leaves only a small region allowed corresponding to positive values of $a/c$ and $-2\lesssim b/c\lesssim-1$.

Interestingly, the allowed regions for both BMs represent isospin violating DM scenarios in which the neutron contribution of the SI component might be much higher than the corresponding proton component but both are generally small in order to evade LUX bounds. For the SD component in BM1, the values of $a_n/a_p$ are not restricted  while for BM1a, the allowed region encodes $a_n/a_p$ generally larger than one. In conclusion, there exists an outstanding complementarity between LHC and direct detection searches for these BM points. While LUX is more stringent than the LHC for negative values of $b/c$, the LHC is more constraining for positive ones, and for BM1a also for negative $a/c$, which highlights the power of combining different experiments in the search for new physics. 

\begin{figure}[tpb]
\centering
\scalebox{0.9}{
            \includegraphics[scale=0.44]{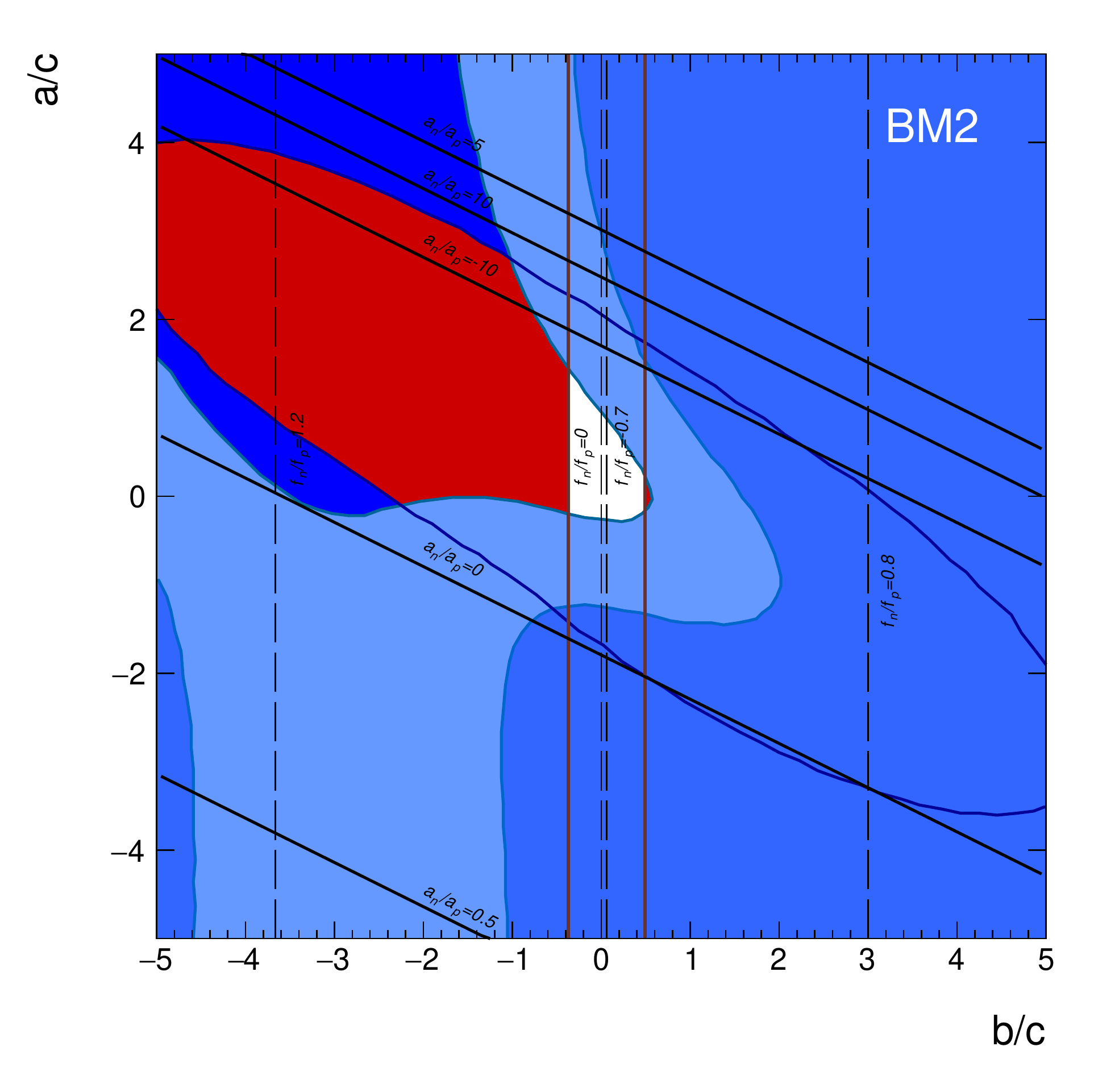}
                        \includegraphics[scale=0.44]{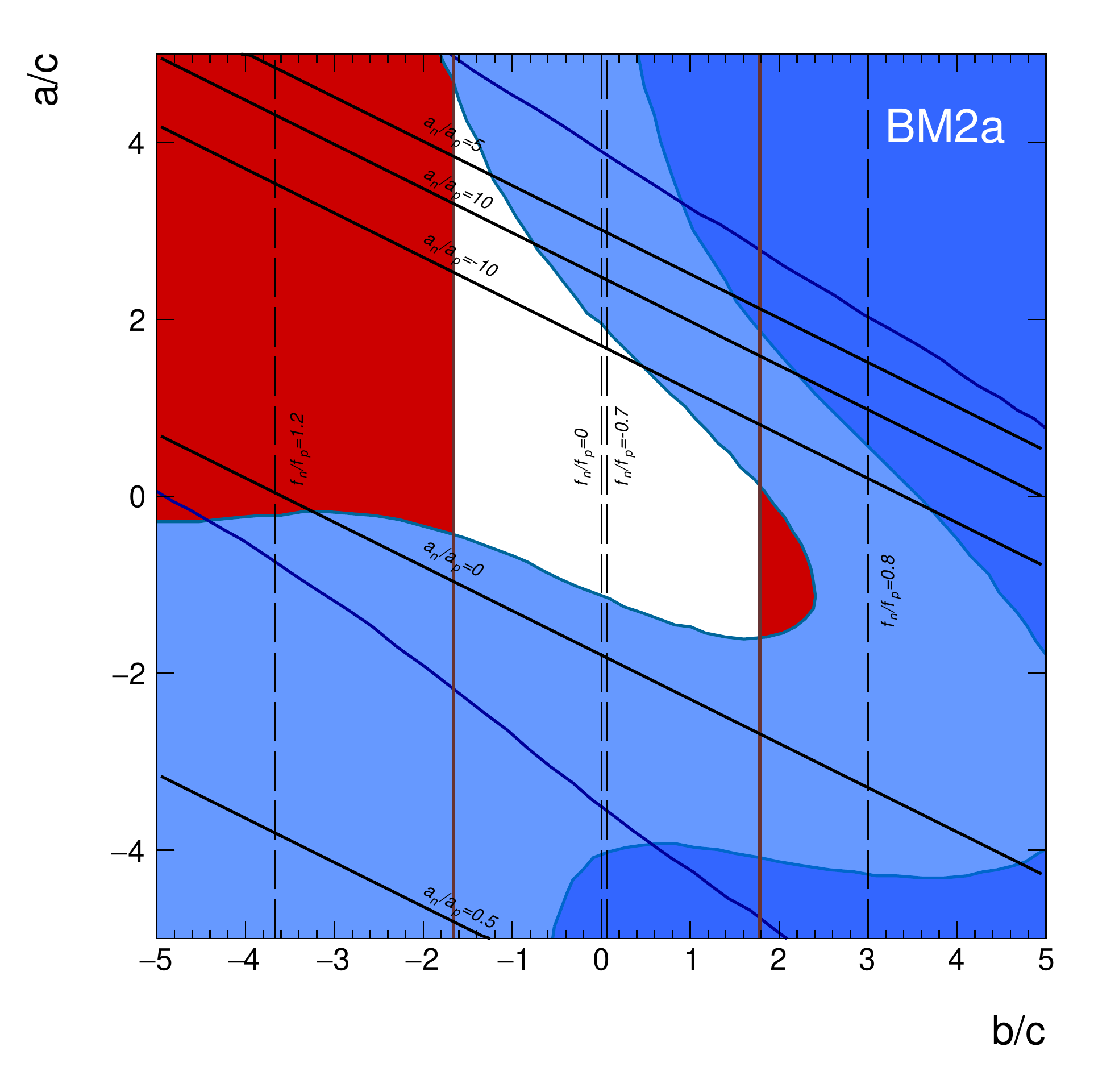}
               }\caption{\small Same as figure~\ref{fig:BM1} but for BM2 (left) and BM2a (right). In this case the exclusion region from dijet resonances at the LHC is shown in dark blue.}\label{fig:BM2}
\end{figure}

Let us move now to BM2 and BM2a. These BM points entail a much heavier DM candidate with respect to the previous ones, now $m_{\psi}=500$~GeV, and a $Z'$ boson of 3 TeV, heavier than before as well. In this region of DM masses, direct detection experiments start to lose their sensitivity very rapidly, so we have increased the DM coupling $h$ in order for the LUX limit to play a role. Besides, by augmenting $d/c$ we have increased the decay width of the $Z'$ boson into leptons, which makes dilepton constraints more stringent. In figure~\ref{fig:BM2} the plane $a/c$-$b/c$ is depicted for these BM points. Notice that in this case dijet searches at the LHC are shown as dark blue regions with oval-like shapes and are specially important in the upper left corner of the BM2 case. 

As in the previous BMs, LUX limits are very stringent in this case, specially for BM2, and again are independent of $a/c$ (dominated by SI interactions). LUX rules out the zone $|b/c|\gtrsim0.4$ for BM2 and $|b/c|\gtrsim1.7$ for BM2a, since for the latter the values of both $c$ and $h$ are smaller. For BM2, the reason for this behaviour is the same as before: in the region not excluded, although the neutron contribution is much higher than the proton contribution, both cross sections are small. LHC limits from dilepton resonances are now very well differentiated and more stringent as a consequence of the increase of $d/c$ (respect to BM1 and BM1a). The difference between $e^+e^-$ and $\mu^+\mu^-$ channels is more notably and comes from the different sensitivity of the ATLAS detector to these channels at this $Z'$ mass, since its coupling to each of these leptons is identical. Finally, dijet resonance searches appear in these cases as more constraining than dileptons and LUX in a small region of the parameter space (the upper left corner in the left panel of figure~\ref{fig:BM2}). The shape of this constraint is due to the squares of the couplings to quarks, involved either in the production mechanism or in the subsequent decay of the $Z'$. This can be understood as a \textit{leptophobic} behaviour of the $Z'$ in this region of BM2, while we have not found such feature in BM2a due to the increase of $d/c$ which makes the $Z'$ more \textit{leptophilic}.

To end with these BM points, as it is shown in Figure~\ref{fig:BM2}, there is only a tiny region allowed for BM2, while for BM2a the region is considerably bigger. In terms of isospin violation in the SI interactions, it corresponds to neutron dominance as in the previous cases. Remarkably, the Xe-phobic scenario ($f_n/f_p=-0.7$) remains allowed by both LHC and LUX in the two BMs analysed. For the SD interactions, the ratio $a_n/a_p$ is found to range between 1 and -10, approximately, and thus, it can be concluded that in general all interactions in direct detection experiments would be dominated by neutrons. The complementarity between direct DM searches and the LHC now takes a new shape. LUX rules out the values of $b/c$ stronger than LHC in all cases, however, the LHC is able to constrain high values of $|a/c|$. Surprisingly, this complementarity is able to delimit the allowed portions of the parameter space so strongly that the we have obtained closed regions. 

\begin{figure}[tpb]
\centering
\scalebox{0.9}{
            \includegraphics[scale=0.44]{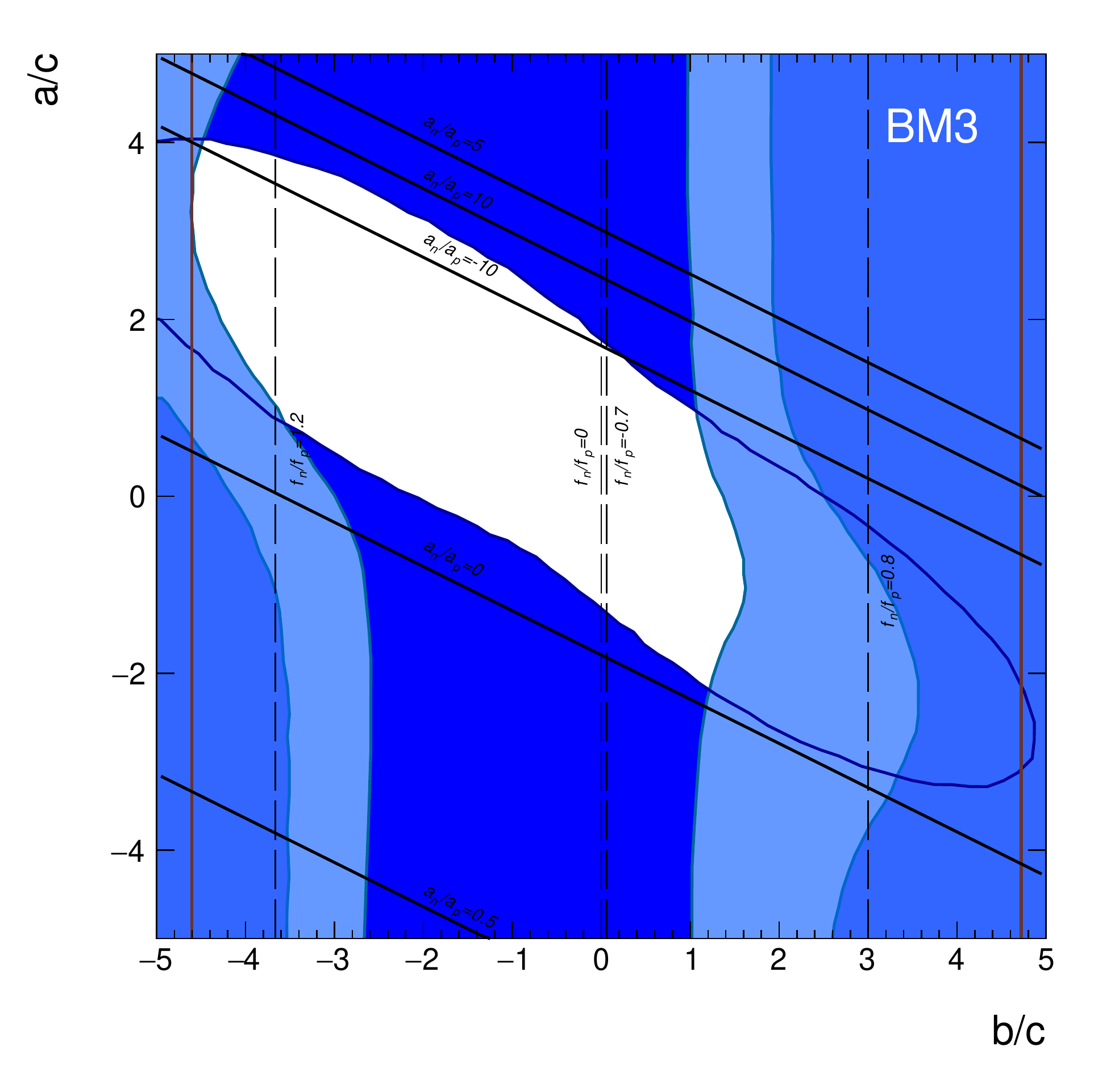}
                        \includegraphics[scale=0.44]{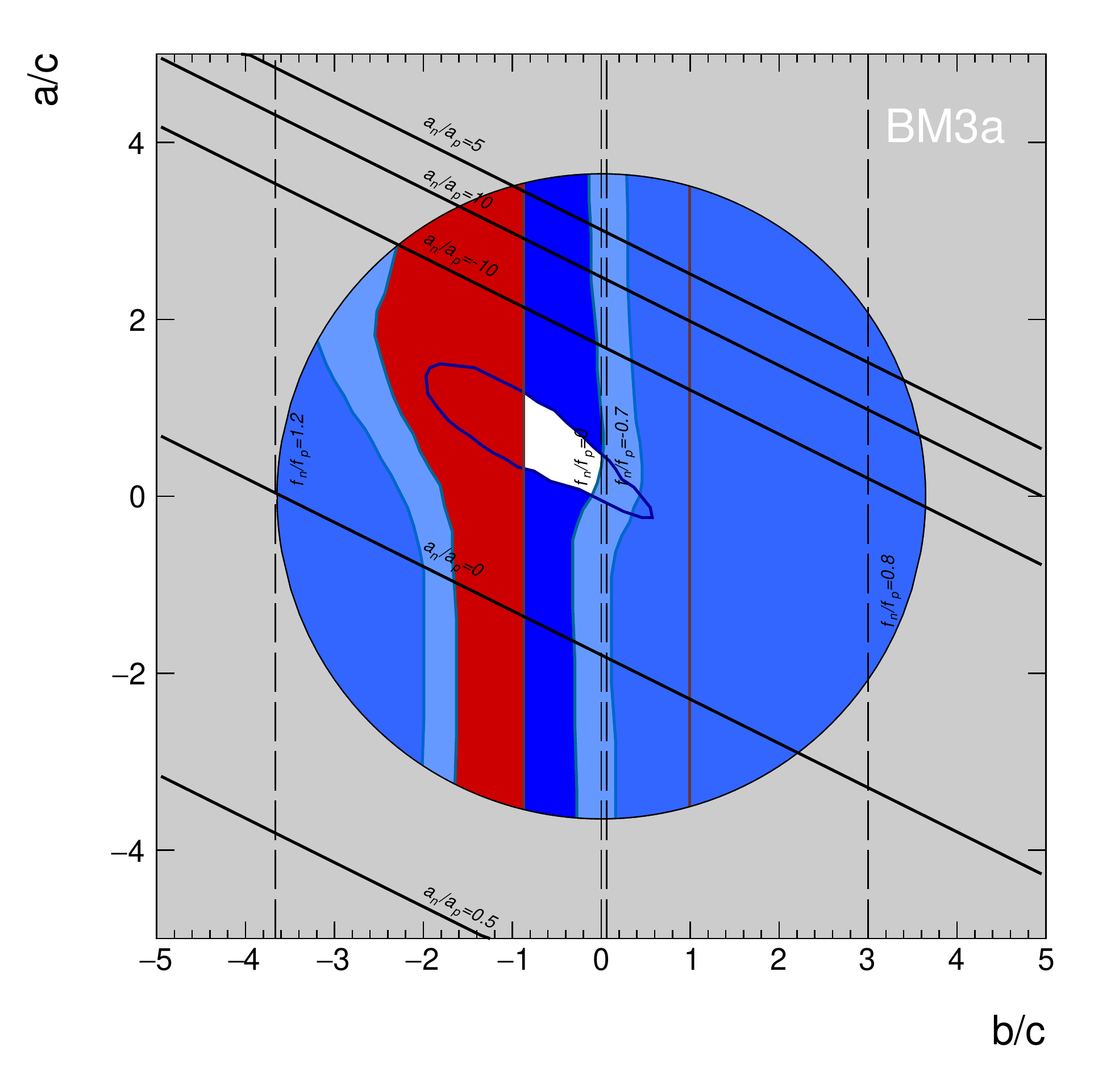}
               }\caption{\small Same as figure~\ref{fig:BM2} but for BM3 (left) and BM3a (right). The grey region on the right panel do not satisfy Eq.~\eqref{eq:norm_visible} and thus, its is not phenomenologically viable.}\label{fig:BM3}
\end{figure}

To end with the analysis, we study two BMs in which the decay of the $Z'$ into DM particles is kinematically forbidden, BM3 and BM3a, unlike for BM1(a) and BM2(a). Our results are shown in figure~\ref{fig:BM3} for BM3 (left panel) and for BM3a (right panel). The choice of the parameters is such that for BM3, LUX limits are not very constraining, while for BM3a the increase of $c$ and $h$ makes LUX very restrictive. However, for the latter a new constraint, very strong, has appeared. The grey area denotes a forbidden region because it does not satisfy Eq.~\eqref{eq:norm_visible}. This is a consequence of the value of $c$ in this case, which is the bigger of all BMs.

Since the $Z'$ boson cannot decay into DM particles in BM3 and BM3a, the branching ratios into SM particles are increased, and therefore, we expect LHC limits to constrain very severly. Notably, for BM3 dijet bounds dominate the region $-3\lesssim b/c\lesssim-1$. The value used for $d/c$ in these BMs makes that for $b/c$ relatively small the $Z'$ boson behaves as \textit{leptophobic}, which results in a decrease in sensitivity of the dilepton searches. As soon as $|b/c|$ increases this behaviour disappears and dilepton bounds are dominant over the dijet ones. In most of the region allowed the SI elastic scattering cross section is dominated by neutrons, except for the region close to $b/c\approx1$. The ratio $a_n/a_p$ allowed is very similar to those in the previous BM points. 

For BM3a, shown in the right panel of figure~\ref{fig:BM3}, we find that only a very small region is allowed. The region extending from $b/c\approx-1$ up to  $b/c\approx0$, and from $a/c\approx0$ to  $a/c\approx1$. From a point of view of complementarity, this region is exceptionally exemplifying since it is delimited by all the searches. The upper and lower regions are bounded by dijet searches, the left by LUX and the right by dilepton searches. This is a consequence of increasing $c$ while keeping the ratio $d/c$ constant. In this case, the SI cross section is dominated by neutrons and the SD proton cross section is similar to the neutron component but with $a_n/a_p\approx-1$.

\section{Conclusions}\label{sec:conclusions}

In this article, we have performed a thorough study of phenomenological features of hidden sector scenarios with St\"uckelberg $Z'$ portals that arise as low energy effective actions of certain type II string compactifications with intersecting branes. For our purposes, the crucial property of these constructions is the unavoidable extension of the SM gauge group by several (`anomalous') abelian gauge bosons which gain a mass and can mix with analogous bosons from hidden sectors.

Many interesting phenomenological properties of such setups are determined by the charges of the SM spectrum under the  extra $U(1)$s of the visible sector, together with a handful of mixing parameters $(a,b,c,d)$. The possible choices for the charges are rather scarce, due to the necessary identification of these symmetries with approximate global symmetries of the SM. We have focussed on a particular gauge structure, the {\it Madrid models} that arises in a large class of intersecting brane constructions. Some other configurations are possible, and they could be studied in analogy. We believe, nevertheless, that our analysis covers a significant portion of the landscape of semi-realistic brane models.

Once the extra visible $U(1)$ bosons mix with those from the hidden sectors, the lightest $Z'$ mass eigenstate generates the dominant interactions between DM and SM fermions.
A particularly appealing and characteristic feature of such models, is the natural appearance of rich patterns of isospin violating DM interactions, which contrasts with other simple portals traditionally considered in the literature. We have explored the prospects for $f_n/f_p$ and $a_n/a_p$ in six different BM points of the parameter space of these constructions, incorporating LHC and LUX bounds showing that in general values of these ratios tend to be dominated by the neutron contribution. Target materials with more sensitivity to neutron interactions are thus very suitable to explore these scenarios. 

Generically, this setup provides isospin violating couplings both in the SI and SD interactions. We have confronted our prospects with LUX and LHC bounds for a set of BM points. By using our own simulation of the LUX experiment, we have performed a check of the exclusion regions for each point using the maximum gap method. This has allowed us to analyse consistently a general scenario with SI and SD (proton and neutron contributions) interactions as well as in general cases of isospin violating couplings of DM. For the LHC we have calculated, for each point of the parameter space, the production cross section of a $Z'$ boson times the branching ratio of a specific decay. With this, we have included ATLAS searches for dilepton ($e^{+}e^{-}$ and $\mu^{+}\mu^{-}$) and dijet resonances. Remarkably, all regions experimentally allowed entail much higher neutron than proton cross sections for the SI interactions while for the SD the situation is less constrained.

The findings of this work open the door to generic scenarios in which the signals in direct detection experiments can be dominated by neutrons. Moreover, we show that the existing complementarity between LHC searches and direct detection experiments is specially relevant to disentangle the couplings of the $Z'$ boson to SM particles. It is gratifying to see how, not only different experimental strategies, but also phenomenological and fundamental theoretical input can be combined into a single framework to shed some light into the possible properties of the so far elusive nature of dark matter.

\subsection*{Acknowledgments}

The authors are grateful to D. G. Cerde\~{n}o, L. Iba\~{n}ez, F. Kahlhoefer and G.~Shiu for useful comments. V.M.L. and M.P. would like to thank the support of the European Union under the ERC Advanced Grant SPLE under contract ERC-2012-ADG-20120216-320421, the support of the Consolider-Ingenio 2010 programme under grant MULTIDARK CSD2009-00064, the Spanish MICINN under Grant No. FPA2012-34694, the Spanish MINECO ``Centro de excelencia Severo Ochoa Program" under Grant No. SEV-2012-0249, and the Community of Madrid under Grant No. HEPHACOS S2009/ESP-1473. P.S. would like to thank DESY, the University of Hamburg, and the Hong Kong IAS for kind hospitality during the completion of this work. He acknowledges  support from the DOE grant DE-FG-02-95ER40896 and the HKRGC grant HKUST4/CRF/13G, 604231, as well as the Collaborative Research Center SFB676 of the DFG at the University of Hamburg.

\bibliographystyle{ieeetr}    
\bibliography{isospin}

\begin{thebibliography}{10}

\bibitem{Patt:2006fw}
B.~Patt and F.~Wilczek, ``{Higgs-field portal into hidden sectors},'' 2006.

\bibitem{Feldman:2007wj}
D.~Feldman, Z.~Liu, and P.~Nath, ``{The Stueckelberg Z-prime Extension with
  Kinetic Mixing and Milli-Charged Dark Matter From the Hidden Sector},'' {\em
  Phys.Rev.}, vol.~D75, p.~115001, 2007.

\bibitem{Falkowski:2009yz}
A.~Falkowski, J.~Juknevich, and J.~Shelton, ``{Dark Matter Through the Neutrino
  Portal},'' 2009.

\bibitem{Batell:2009di}
B.~Batell, M.~Pospelov, and A.~Ritz, ``{Exploring Portals to a Hidden Sector
  Through Fixed Targets},'' {\em Phys.Rev.}, vol.~D80, p.~095024, 2009.

\bibitem{Crivelli:2010bk}
P.~Crivelli, A.~Belov, U.~Gendotti, S.~Gninenko, and A.~Rubbia, ``{Positronium
  Portal into Hidden Sector: A new Experiment to Search for Mirror Dark
  Matter},'' {\em JINST}, vol.~5, p.~P08001, 2010.

\bibitem{Chu:2011be}
X.~Chu, T.~Hambye, and M.~H. Tytgat, ``{The Four Basic Ways of Creating Dark
  Matter Through a Portal},'' {\em JCAP}, vol.~1205, p.~034, 2012.

\bibitem{Essig:2013lka}
R.~Essig, J.~A. Jaros, W.~Wester, P.~H. Adrian, S.~Andreas, {\em et~al.},
  ``{Working Group Report: New Light Weakly Coupled Particles},'' 2013.

\bibitem{Feng:2014cla}
W.-Z. Feng, G.~Shiu, P.~Soler, and F.~Ye, ``{Building a Stückelberg portal},''
  {\em JHEP}, vol.~1405, p.~065, 2014.

\bibitem{Feng:2014eja}
W.-Z. Feng, G.~Shiu, P.~Soler, and F.~Ye, ``{Probing Hidden Sectors with
  St\"uckelberg U(1) Gauge Fields},'' {\em Phys.Rev.Lett.}, vol.~113,
  p.~061802, 2014.

\bibitem{Foot:2014uba}
R.~Foot and S.~Vagnozzi, ``{Dissipative hidden sector dark matter},'' {\em
  Phys.Rev.}, vol.~D91, no.~2, p.~023512, 2015.

\bibitem{Bai:2014osa}
Y.~Bai and J.~Berger, ``{Lepton Portal Dark Matter},'' {\em JHEP}, vol.~1408,
  p.~153, 2014.

\bibitem{Baek:2013dwa}
S.~Baek, P.~Ko, and W.-I. Park, ``{Hidden sector monopole, vector dark matter
  and dark radiation with Higgs portal},'' {\em JCAP}, vol.~1410, no.~10,
  p.~067, 2014.

\bibitem{Blum:2014jca}
K.~Blum, M.~Cliche, C.~Csaki, and S.~J. Lee, ``{WIMP Dark Matter through the
  Dilaton Portal},'' 2014.

\bibitem{Cherry:2014xra}
J.~F. Cherry, A.~Friedland, and I.~M. Shoemaker, ``{Neutrino Portal Dark
  Matter: From Dwarf Galaxies to IceCube},'' 2014.

\bibitem{Arcadi:2014lta}
G.~Arcadi, Y.~Mambrini, and F.~Richard, ``{Z-portal dark matter},'' 2014.

\bibitem{Bian:2014cja}
L.~Bian, T.~Li, J.~Shu, and X.-C. Wang, ``{Two component dark matter with
  multi-Higgs portals},'' 2014.

\bibitem{Djouadi:2011aa}
A.~Djouadi, O.~Lebedev, Y.~Mambrini, and J.~Quevillon, ``{Implications of LHC
  searches for Higgs--portal dark matter},'' {\em Phys.Lett.}, vol.~B709,
  pp.~65--69, 2012.

\bibitem{Djouadi:2012zc}
A.~Djouadi, A.~Falkowski, Y.~Mambrini, and J.~Quevillon, ``{Direct Detection of
  Higgs-Portal Dark Matter at the LHC},'' {\em Eur.Phys.J.}, vol.~C73, no.~6,
  p.~2455, 2013.

\bibitem{Langacker:2008yv}
P.~Langacker, ``{The Physics of Heavy $Z^\prime$ Gauge Bosons},'' {\em
  Rev.Mod.Phys.}, vol.~81, pp.~1199--1228, 2009.

\bibitem{Dudas:2009uq}
E.~Dudas, Y.~Mambrini, S.~Pokorski, and A.~Romagnoni, ``{(In)visible Z-prime
  and dark matter},'' {\em JHEP}, vol.~0908, p.~014, 2009.

\bibitem{Cassel:2009pu}
S.~Cassel, D.~Ghilencea, and G.~Ross, ``{Electroweak and Dark Matter
  Constraints on a Z-prime in Models with a Hidden Valley},'' {\em Nucl.Phys.},
  vol.~B827, pp.~256--280, 2010.

\bibitem{Frandsen:2011cg}
M.~T. Frandsen, F.~Kahlhoefer, S.~Sarkar, and K.~Schmidt-Hoberg, ``{Direct
  detection of dark matter in models with a light Z'},'' {\em JHEP}, vol.~1109,
  p.~128, 2011.

\bibitem{Barger:2012ey}
V.~Barger, D.~Marfatia, and A.~Peterson, ``{LHC and dark matter signals of Z′
  bosons},'' {\em Phys.Rev.}, vol.~D87, no.~1, p.~015026, 2013.

\bibitem{Arcadi:2013qia}
G.~Arcadi, Y.~Mambrini, M.~H.~G. Tytgat, and B.~Zaldivar, ``{Invisible
  $Z^\prime$ and dark matter: LHC vs LUX constraints},'' {\em JHEP}, vol.~1403,
  p.~134, 2014.

\bibitem{Alves:2013tqa}
A.~Alves, S.~Profumo, and F.~S. Queiroz, ``{The dark $Z^{'}$ portal: direct,
  indirect and collider searches},'' {\em JHEP}, vol.~1404, p.~063, 2014.

\bibitem{Alves:2015pea}
A.~Alves, A.~Berlin, S.~Profumo, and F.~S. Queiroz, ``{Dark Matter
  Complementarity and the Z$^\prime$ Portal},'' 2015.

\bibitem{Gao:2011ka}
X.~Gao, Z.~Kang, and T.~Li, ``{Origins of the Isospin Violation of Dark Matter
  Interactions},'' {\em JCAP}, vol.~1301, p.~021, 2013.

\bibitem{Belanger:2013tla}
G.~B\'{e}langer, A.~Goudelis, J.-C. Park, and A.~Pukhov, ``{Isospin-violating dark
  matter from a double portal},'' {\em JCAP}, vol.~1402, p.~020, 2014.

\bibitem{Hamaguchi:2014pja}
K.~Hamaguchi, S.~P. Liew, T.~Moroi, and Y.~Yamamoto, ``{Isospin-Violating Dark
  Matter with Colored Mediators},'' {\em JHEP}, vol.~1405, p.~086, 2014.

\bibitem{Ibanez:2012zz}
L.~E. Ib\'a\~nez and A.~M. Uranga, ``{String theory and particle physics: An
  introduction to string phenomenology},'' 2012.

\bibitem{Blumenhagen:2005mu}
R.~Blumenhagen, M.~Cvetic, P.~Langacker, and G.~Shiu, ``{Toward realistic
  intersecting D-brane models},'' {\em Ann.Rev.Nucl.Part.Sci.}, vol.~55,
  pp.~71--139, 2005.

\bibitem{Blumenhagen:2006ci}
R.~Blumenhagen, B.~Kors, D.~Lust, and S.~Stieberger, ``{Four-dimensional String
  Compactifications with D-Branes, Orientifolds and Fluxes},'' {\em
  Phys.Rept.}, vol.~445, pp.~1--193, 2007.

\bibitem{Marchesano:2007de}
F.~Marchesano, ``{Progress in D-brane model building},'' {\em Fortsch.Phys.},
  vol.~55, pp.~491--518, 2007.

\bibitem{Kakushadze:1997mc}
Z.~Kakushadze, G.~Shiu, S.~H. Tye, and Y.~Vtorov-Karevsky, ``{A Review of three
  family grand unified string models},'' {\em Int.J.Mod.Phys.}, vol.~A13,
  pp.~2551--2598, 1998.

\bibitem{Cleaver:1998gc}
G.~Cleaver, M.~Cvetic, J.~Espinosa, L.~Everett, P.~Langacker, {\em et~al.},
  ``{Physics implications of flat directions in free fermionic superstring
  models 1. Mass spectrum and couplings},'' {\em Phys.Rev.}, vol.~D59,
  p.~055005, 1999.

\bibitem{Holdom:1985ag}
B.~Holdom, ``{Two U(1)'s and Epsilon Charge Shifts},'' {\em Phys.Lett.},
  vol.~B166, p.~196, 1986.

\bibitem{Lust:2003ky}
D.~Lust and S.~Stieberger, ``{Gauge threshold corrections in intersecting brane
  world models},'' {\em Fortsch.Phys.}, vol.~55, pp.~427--465, 2007.

\bibitem{Abel:2003ue}
S.~Abel and B.~Schofield, ``{Brane anti-brane kinetic mixing, millicharged
  particles and SUSY breaking},'' {\em Nucl.Phys.}, vol.~B685, pp.~150--170,
  2004.

\bibitem{Abel:2006qt}
S.~A. Abel, J.~Jaeckel, V.~V. Khoze, and A.~Ringwald, ``{Illuminating the
  Hidden Sector of String Theory by Shining Light through a Magnetic Field},''
  {\em Phys.Lett.}, vol.~B666, pp.~66--70, 2008.

\bibitem{Abel:2008ai}
S.~Abel, M.~Goodsell, J.~Jaeckel, V.~Khoze, and A.~Ringwald, ``{Kinetic Mixing
  of the Photon with Hidden U(1)s in String Phenomenology},'' {\em JHEP},
  vol.~0807, p.~124, 2008.

\bibitem{Goodsell:2009xc}
M.~Goodsell, J.~Jaeckel, J.~Redondo, and A.~Ringwald, ``{Naturally Light Hidden
  Photons in LARGE Volume String Compactifications},'' {\em JHEP}, vol.~0911,
  p.~027, 2009.

\bibitem{Cicoli:2011yh}
M.~Cicoli, M.~Goodsell, J.~Jaeckel, and A.~Ringwald, ``{Testing String Vacua in
  the Lab: From a Hidden CMB to Dark Forces in Flux Compactifications},'' {\em
  JHEP}, vol.~1107, p.~114, 2011.

\bibitem{Gmeiner:2009fb}
F.~Gmeiner and G.~Honecker, ``{Complete Gauge Threshold Corrections for
  Intersecting Fractional D6-Branes: The Z6 and Z6' Standard Models},'' {\em
  Nucl.Phys.}, vol.~B829, pp.~225--297, 2010.

\bibitem{Honecker:2011sm}
G.~Honecker, ``{Kaehler metrics and gauge kinetic functions for intersecting
  D6-branes on toroidal orbifolds - The complete perturbative story},'' {\em
  Fortsch.Phys.}, vol.~60, pp.~243--326, 2012.

\bibitem{Kors:2004dx}
B.~Kors and P.~Nath, ``{A Stueckelberg extension of the standard model},'' {\em
  Phys.Lett.}, vol.~B586, pp.~366--372, 2004.

\bibitem{Dermisek:2007qi}
R.~Dermisek, H.~Verlinde, and L.-T. Wang, ``{Hypercharged Anomaly Mediation},''
  {\em Phys.Rev.Lett.}, vol.~100, p.~131804, 2008.

\bibitem{Verlinde:2007qk}
H.~Verlinde, L.-T. Wang, M.~Wijnholt, and I.~Yavin, ``{A Higher Form (of)
  Mediation},'' {\em JHEP}, vol.~0802, p.~082, 2008.

\bibitem{Shiu:2013wxa}
G.~Shiu, P.~Soler, and F.~Ye, ``{Millicharged Dark Matter in Quantum Gravity
  and String Theory},'' {\em Phys.Rev.Lett.}, vol.~110, no.~24, p.~241304,
  2013.

\bibitem{Ghilencea:2002da}
D.~Ghilencea, L.~Ib\'a\~nez, N.~Irges, and F.~Quevedo, ``{TeV scale Z-prime bosons
  from D-branes},'' {\em JHEP}, vol.~0208, p.~016, 2002.

\bibitem{Blumenhagen:2006xt}
R.~Blumenhagen, M.~Cvetic, and T.~Weigand, ``{Spacetime instanton corrections
  in 4D string vacua: The Seesaw mechanism for D-Brane models},'' {\em
  Nucl.Phys.}, vol.~B771, pp.~113--142, 2007.

\bibitem{Ibanez:2006da}
L.~Ib\'a\~nez and A.~Uranga, ``{Neutrino Majorana Masses from String Theory
  Instanton Effects},'' {\em JHEP}, vol.~0703, p.~052, 2007.

\bibitem{Florea:2006si}
B.~Florea, S.~Kachru, J.~McGreevy, and N.~Saulina, ``{Stringy Instantons and
  Quiver Gauge Theories},'' {\em JHEP}, vol.~0705, p.~024, 2007.

\bibitem{Ibanez:2001nd}
L.~E. Ib\'a\~nez, F.~Marchesano, and R.~Rabadan, ``{Getting just the standard model
  at intersecting branes},'' {\em JHEP}, vol.~0111, p.~002, 2001.

\bibitem{Feng:2011vu}
J.~L. Feng, J.~Kumar, D.~Marfatia, and D.~Sanford, ``{Isospin-Violating Dark
  Matter},'' {\em Phys.Lett.}, vol.~B703, pp.~124--127, 2011.

\bibitem{Chun:2010ve}
E.~J. Chun, J.-C. Park, and S.~Scopel, ``{Dark matter and a new gauge boson
  through kinetic mixing},'' {\em JHEP}, vol.~1102, p.~100, 2011.

\bibitem{Drozd:2014yla}
A.~Drozd, B.~Grzadkowski, J.~F. Gunion, and Y.~Jiang, ``{Extending
  two-Higgs-doublet models by a singlet scalar field - the Case for Dark
  Matter},'' {\em JHEP}, vol.~1411, p.~105, 2014.

\bibitem{Cremades:2003qj}
D.~Cremades, L.~Ib\'a\~nez, and F.~Marchesano, ``{Yukawa couplings in intersecting
  D-brane models},'' {\em JHEP}, vol.~0307, p.~038, 2003.

\bibitem{Cerdeno:2010jj}
D.~G. Cerde\~no and A.~M. Green, ``{Direct detection of WIMPs},'' 2010.

\bibitem{Belanger:2013oya}
G.~B\'elanger, F.~Boudjema, A.~Pukhov, and A.~Semenov, ``{micrOMEGAs3: A program
  for calculating dark matter observables},'' {\em Comput.Phys.Commun.},
  vol.~185, pp.~960--985, 2014.

\bibitem{Klos:2013rwa}
P.~Klos, J.~Men\'endez, D.~Gazit, and A.~Schwenk, ``{Large-scale nuclear
  structure calculations for spin-dependent WIMP scattering with chiral
  effective field theory currents},'' {\em Phys.Rev.}, vol.~D88, no.~8,
  p.~083516, 2013.

\bibitem{Fayet:2007ua}
P.~Fayet, ``{U-boson production in e+ e- annihilations, psi and Upsilon decays,
  and Light Dark Matter},'' {\em Phys.Rev.}, vol.~D75, p.~115017, 2007.

\bibitem{Teubner:2010ah}
T.~Teubner, K.~Hagiwara, R.~Liao, A.~Martin, and D.~Nomura, ``{Update of g-2 of
  the Muon and Delta Alpha},'' {\em Chin.Phys.}, vol.~C34, pp.~728--734, 2010.

\bibitem{Akerib:2013tjd}
D.~Akerib {\em et~al.}, ``{First results from the LUX dark matter experiment at
  the Sanford Underground Research Facility},'' {\em Phys.Rev.Lett.}, vol.~112,
  p.~091303, 2014.

\bibitem{Yellin:2002xd}
S.~Yellin, ``{Finding an upper limit in the presence of unknown background},''
  {\em Phys.Rev.}, vol.~D66, p.~032005, 2002.

\bibitem{Aprile:2011hx}
E.~Aprile {\em et~al.}, ``{Likelihood Approach to the First Dark Matter Results
  from XENON100},'' {\em Phys.Rev.}, vol.~D84, p.~052003, 2011.

\bibitem{pmt}
D.~Akerib {\em et~al.}, ``{The Large Underground Xenon (LUX) Experiment},''
  {\em Nucl.Instrum.Meth.}, vol.~A704, pp.~111--126, 2013.

\bibitem{Cerdeno:2012ix}
D.~Cerde\~no, M.~Fornasa, J.-H. Huh, and M.~Peir\'o, ``{Nuclear uncertainties in
  the spin-dependent structure functions for direct dark matter detection},''
  {\em Phys.Rev.}, vol.~D87, no.~2, p.~023512, 2013.

\bibitem{Aad:2014cka}
G.~Aad {\em et~al.}, ``{Search for high-mass dilepton resonances in pp
  collisions at $\sqrt{s}=8$  TeV with the ATLAS detector},'' {\em
  Phys.Rev.}, vol.~D90, no.~5, p.~052005, 2014.

\bibitem{Aaltonen:2008dn}
T.~Aaltonen {\em et~al.}, ``{Search for new particles decaying into dijets in
  proton-antiproton collisions at s**(1/2) = 1.96-TeV},'' {\em Phys.Rev.},
  vol.~D79, p.~112002, 2009.

\bibitem{Aad:2011fq}
G.~Aad {\em et~al.}, ``{Search for New Physics in the Dijet Mass Distribution
  using 1 fb$^{-1}$ of $pp$ Collision Data at $\sqrt{s}=7$ TeV collected by the
  ATLAS Detector},'' {\em Phys.Lett.}, vol.~B708, pp.~37--54, 2012.

\bibitem{CMS-dijet}
G.~Aad {\em et~al.}, ``{The CMS Collaboration, Search for Narrow Resonances
  using the Dijet Mass Spectrum with 19.6 fb$^{-1}$ of $pp$ Collisions at
  $\sqrt{s} = 8$~TeV, CMS-PAS-EXO-12-059 (2013)},''

\bibitem{Carena:2004xs}
M.~S. Carena, A.~Daleo, B.~A. Dobrescu, and T.~M. Tait, ``{$Z^\prime$ gauge
  bosons at the Tevatron},'' {\em Phys.Rev.}, vol.~D70, p.~093009, 2004.

\bibitem{Belyaev:2012qa}
A.~Belyaev, N.~D. Christensen, and A.~Pukhov, ``{CalcHEP 3.4 for collider
  physics within and beyond the Standard Model},'' {\em Comput.Phys.Commun.},
  vol.~184, pp.~1729--1769, 2013.

\bibitem{Accomando:2010fz}
E.~Accomando, A.~Belyaev, L.~Fedeli, S.~F. King, and
  C.~Shepherd-Themistocleous, ``{Z' physics with early LHC data},'' {\em
  Phys.Rev.}, vol.~D83, p.~075012, 2011.

\end{thebibliography}

\end{document}